\documentclass[lettersize,journal]{IEEEtran}
\IEEEoverridecommandlockouts
\usepackage{cite}
\usepackage[T1]{fontenc}
\usepackage{graphicx}
\usepackage{amssymb}
\usepackage{amsmath}
\usepackage{amsthm}
\usepackage{subfigure}
\usepackage{microtype}
\usepackage{balance}
\usepackage{xcolor}
\usepackage{algorithm}
\usepackage{algorithmicx}
\usepackage{algpseudocode}
\usepackage[acronym]{glossaries}
\usepackage{url}
\usepackage{float}
\usepackage{booktabs}
\usepackage{hyperref}
\usepackage{graphicx}

\algrenewcommand\algorithmicindent{0.7em}%

\newacronym{NPRACH}{NPRACH}{narrowband physical random-access channel}
\newacronym{ToA}{ToA}{time of arrival}
\newacronym{CFO}{CFO}{carrier frequency offset}
\newacronym{NBIoT}{NB-IoT}{narrowband internet of things}
\newacronym{5GNR}{5G NR}{5G New Radio}
\newacronym{3GPP}{3GPP}{3rd Generation Partnership Project}
\newacronym{UMi}{UMi}{urban microcell}
\newacronym{RMSE}{RMSE}{root-mean-square error}
\newacronym{NN}{NN}{neural network}
\newacronym{BS}{BS}{base station}
\newacronym{UE}{UE}{user equipment}
\newacronym{SG}{SG}{symbol group}
\newacronym{CP}{CP}{cyclic prefix}
\newacronym{OFDM}{OFDM}{orthogonal frequency division multiplexing}
\newacronym{FFT}{FFT}{fast Fourier transform}
\newacronym{AWGN}{AWGN}{additive white Gaussian noise}
\newacronym{DFT}{DFT}{discrete Fourier transform}
\newacronym{FNR}{FNR}{false negative rate}
\newacronym{FPR}{FPR}{false positive rate}
\newacronym{RG}{RG}{resource grid}
\newacronym{RE}{RE}{resource element}
\newacronym{SNR}{SNR}{signal-to-noise ratio}
\newacronym{1D}{1D}{one-dimensional}
\newacronym{MLP}{MLP}{multilayer perceptron}
\newacronym{BCE}{BCE}{binary cross-entropy}
\newacronym{KL}{KL}{Kullback–Leibler}
\newacronym{SGD}{SGD}{stochastic gradient descent}
\newacronym{ppm}{ppm}{parts-per-million}
\newacronym{ICI}{ICI}{inter-carrier interference}
\newacronym{GNN}{GNN}{graph neural network}
\newacronym{BP}{BP}{belief propagation}
\newacronym{FEC}{FEC}{forward error correction}
\newacronym{LDPC}{LDPC}{low-density parity-check}
\newacronym{HDPC}{HDPC}{high-density parity-check}
\newacronym{SCL}{SCL}{successive cancellation list}
\newacronym{SC}{SC}{successive cancellation}
\newacronym{URLLC}{URLLC}{ultra-reliable low-latency communications}
\newacronym{APP}{APP}{a posterior probability}
\newacronym{MIMO}{MIMO}{multiple-input multiple-output}
\newacronym{CNN}{CNN}{convolutional neural network}
\newacronym{BER}{BER}{bit error rate}
\newacronym{BPSK}{BPSK}{binary phase shift keying}
\newacronym{LLR}{LLR}{log-likelihood ratio}
\newacronym{FN}{FN}{factor node}
\newacronym{VN}{VN}{variable node}
\newacronym{CN}{CN}{check node}
\newacronym{MPNN}{MPNN}{message passing neural network}

\newacronym{AI}{AI}{artificial intelligence}
\newacronym{ML}{ML}{machine learning}
\newacronym{SISO}{SISO}{single input single output}
\newacronym{PRB}{PRB}{physical resource block}
\newacronym{PUSCH}{PUSCH}{physical uplink shared channel}

\newacronym{MUMIMO}{MU-MIMO}{multi-user multiple-input multiple-output}
\newacronym{BICM}{BICM}{bit-interleaved coded modulation}
\newacronym{QAM}{QAM}{quadrature amplitude modulation}
\newacronym{LMMSE}{LMMSE}{linear minimum mean square error}
\newacronym{MMSE}{MMSE}{minimum mean square error}
\newacronym{CSI}{CSI}{channel-state information}
\newacronym{SIMO}{SIMO}{single-input multiple-output}
\newacronym{CGNN}{CGNN}{convolutional and graph neural network}
\newacronym{BLER}{BLER}{block error rate}
\newacronym{LS}{LS}{least squares}
\newacronym{PE}{PE}{positional encoding}
\newacronym{relu}{ReLU}{rectified linear unit}
\newacronym{RB}{RB}{resource block}
\newacronym{CGGNN}{CGGNN}{convolutional graph neural network}
\newacronym{DMRS}{DMRS}{demodulation reference signal}
\newacronym{IoT}{IoT}{internet of things}
\newacronym{ADAM}{ADAM}{adaptive momentum}
\newacronym{TBLER}{TBLER}{transport block error rate}
\newacronym{MCS}{MCS}{modulation and coding scheme}
\newacronym{TDL}{TDL}{tapped delay line}
\newacronym{CDM}{CDM}{code division multiplexing}
\newacronym{FLOP}{FLOP}{floating point operation}
\newacronym{PHY}{PHY}{physical layer}

\newacronym{ULA}{ULA}{uniform linear array}
\newacronym{NRX}{NRX}{neural receiver}
\newacronym{Var-MCS-NRX}{Var-MCS NRX}{variable-MCS NRX}
\newacronym{UL}{UL}{uplink}
\newacronym{DL}{DL}{downlink}
\newacronym{MSE}{MSE}{mean squared error}
\newacronym{CIR}{CIR}{channel impulse response}
\newacronym{iid}{iid}{independent and identically distributed}

\newacronym{RAN}{RAN}{radio access network}
\newacronym{ORU}{O-RU}{open RAN radio unit}
\newacronym{COTS}{COTS}{commercial-off-the-shelf}
\newacronym{RF}{RF}{radio frequency}
\newacronym{LOS}{LOS}{line of sight}
\newacronym{OTA}{OTA}{over-the-air}
\newacronym{TDD}{TDD}{time-division duplexing}
\newacronym{CAEZ}{CAEZ}{CSI acquisition at ETH Zurich}
\newacronym{ARC-OTA}{ARC-OTA}{Aerial RAN CoLab Over-the-Air}
\newacronym{OAI}{OAI}{OpenAirInterface}
\newacronym{RFFI}{RFFI}{radio frequency fingerprint identification}
\newacronym{FH}{FH}{front haul}
\newacronym{GNSS}{GNSS}{global navigation satellite system}
\newacronym{PTP}{PTP}{precision time protocol}
\newacronym{RNTI}{RNTI}{radio network temporary identifier}
\newacronym{ISM}{ISM}{Industrial, Scientific, and Medical}
\usepackage{amssymb}
\usepackage{amsfonts}
\usepackage{mathrsfs}
\usepackage{xspace}
\usepackage{bm}
\usepackage{upgreek}

\newcommand{\safemath}[2]{\newcommand{#1}{\ensuremath{#2}\xspace}}

\safemath{\bma}{\mathbf{a}}
\safemath{\bmb}{\mathbf{b}}
\safemath{\bmc}{\mathbf{c}}
\safemath{\bmd}{\mathbf{d}}
\safemath{\bme}{\mathbf{e}}
\safemath{\bmf}{\mathbf{f}}
\safemath{\bmg}{\mathbf{g}}
\safemath{\bmh}{\mathbf{h}}
\safemath{\bmi}{\mathbf{i}}
\safemath{\bmj}{\mathbf{j}}
\safemath{\bmk}{\mathbf{k}}
\safemath{\bml}{\mathbf{l}}
\safemath{\bmm}{\mathbf{m}}
\safemath{\bmn}{\mathbf{n}}
\safemath{\bmo}{\mathbf{o}}
\safemath{\bmp}{\mathbf{p}}
\safemath{\bmq}{\mathbf{q}}
\safemath{\bmr}{\mathbf{r}}
\safemath{\bms}{\mathbf{s}}
\safemath{\bmt}{\mathbf{t}}
\safemath{\bmu}{\mathbf{u}}
\safemath{\bmv}{\mathbf{v}}
\safemath{\bmw}{\mathbf{w}}
\safemath{\bmx}{\mathbf{x}}
\safemath{\bmy}{\mathbf{y}}
\safemath{\bmz}{\mathbf{z}}
\safemath{\bmzero}{\mathbf{0}}
\safemath{\bmone}{\mathbf{1}}
\safemath{\Bell}{\ensuremath{\boldsymbol\ell}}

\bmdefine{\biad}{a}
\bmdefine{\bibd}{b}
\bmdefine{\bicd}{c}
\bmdefine{\bidd}{d}
\bmdefine{\bied}{e}
\bmdefine{\bifd}{f}
\bmdefine{\bigd}{g}
\bmdefine{\bihd}{h}
\bmdefine{\biid}{i}
\bmdefine{\bijd}{j}
\bmdefine{\bikd}{k}
\bmdefine{\bild}{l}
\bmdefine{\bimd}{m}
\bmdefine{\bind}{n}
\bmdefine{\biod}{o}
\bmdefine{\bipd}{p}
\bmdefine{\biqd}{q}
\bmdefine{\bird}{r}
\bmdefine{\bisd}{s}
\bmdefine{\bitd}{t}
\bmdefine{\biud}{u}
\bmdefine{\bivd}{v}
\bmdefine{\biwd}{w}
\bmdefine{\bixd}{x}
\bmdefine{\biyd}{y}
\bmdefine{\bizd}{z}

\bmdefine{\bixid}{\xi}
\bmdefine{\bilambdad}{\lambda}
\bmdefine{\bimud}{\mu}
\bmdefine{\bithetad}{\theta}
\bmdefine{\biphid}{\phi}
\bmdefine{\bideltad}{\delta}

\safemath{\bmia}{\biad}
\safemath{\bmib}{\bibd}
\safemath{\bmic}{\bicd}
\safemath{\bmid}{\bidd}
\safemath{\bmie}{\bied}
\safemath{\bmif}{\bifd}
\safemath{\bmig}{\bigd}
\safemath{\bmih}{\bihd}
\safemath{\bmii}{\biid}
\safemath{\bmij}{\bijd}
\safemath{\bmik}{\bikd}
\safemath{\bmil}{\bild}
\safemath{\bmim}{\bimd}
\safemath{\bmin}{\bind}
\safemath{\bmio}{\biod}
\safemath{\bmip}{\bipd}
\safemath{\bmiq}{\biqd}
\safemath{\bmir}{\bird}
\safemath{\bmis}{\bisd}
\safemath{\bmit}{\bitd}
\safemath{\bmiu}{\biud}
\safemath{\bmiv}{\bivd}
\safemath{\bmiw}{\biwd}
\safemath{\bmix}{\bixd}
\safemath{\bmiy}{\biyd}
\safemath{\bmiz}{\bizd}

\safemath{\bmxi}{\bixid}
\safemath{\bmlambda}{\bilambdad}
\safemath{\bmmu}{\bimud}
\safemath{\bmtheta}{\bithetad}
\safemath{\bmphi}{\biphid}
\safemath{\bmdelta}{\bideltad}

\safemath{\bA}{\mathbf{A}}
\safemath{\bB}{\mathbf{B}}
\safemath{\bC}{\mathbf{C}}
\safemath{\bD}{\mathbf{D}}
\safemath{\bE}{\mathbf{E}}
\safemath{\bF}{\mathbf{F}}
\safemath{\bG}{\mathbf{G}}
\safemath{\bH}{\mathbf{H}}
\safemath{\bI}{\mathbf{I}}
\safemath{\bJ}{\mathbf{J}}
\safemath{\bK}{\mathbf{K}}
\safemath{\bL}{\mathbf{L}}
\safemath{\bM}{\mathbf{M}}
\safemath{\bN}{\mathbf{N}}
\safemath{\bO}{\mathbf{O}}
\safemath{\bP}{\mathbf{P}}
\safemath{\bQ}{\mathbf{Q}}
\safemath{\bR}{\mathbf{R}}
\safemath{\bS}{\mathbf{S}}
\safemath{\bT}{\mathbf{T}}
\safemath{\bU}{\mathbf{U}}
\safemath{\bV}{\mathbf{V}}
\safemath{\bW}{\mathbf{W}}
\safemath{\bX}{\mathbf{X}}
\safemath{\bY}{\mathbf{Y}}
\safemath{\bZ}{\mathbf{Z}}

\safemath{\bZero}{\mathbf{0}}
\safemath{\bOne}{\mathbf{1}}
\safemath{\bDelta}{\mathbf{\Delta}}
\safemath{\bLambda}{\mathbf{\UpLambda}}
\safemath{\bPhi}{\mathbf{\Upphi}}
\safemath{\bSigma}{\mathbf{\Upsigma}}
\safemath{\bOmega}{\mathbf{\Upomega}}
\safemath{\bTheta}{\mathbf{\Uptheta}}

\bmdefine{\biAd}{A}
\bmdefine{\biBd}{B}
\bmdefine{\biCd}{C}
\bmdefine{\biDd}{D}
\bmdefine{\biEd}{E}
\bmdefine{\biFd}{F}
\bmdefine{\biGd}{G}
\bmdefine{\biHd}{H}
\bmdefine{\biId}{I}
\bmdefine{\biJd}{J}
\bmdefine{\biKd}{K}
\bmdefine{\biLd}{L}
\bmdefine{\biMd}{M}
\bmdefine{\biOd}{N}
\bmdefine{\biPd}{O}
\bmdefine{\biQd}{P}
\bmdefine{\biRd}{R}
\bmdefine{\biSd}{S}
\bmdefine{\biTd}{T}
\bmdefine{\biUd}{U}
\bmdefine{\biVd}{V}
\bmdefine{\biWd}{W}
\bmdefine{\biXd}{X}
\bmdefine{\biYd}{Y}
\bmdefine{\biZd}{Z}

\bmdefine{\biDelta}{\Delta}
\bmdefine{\biLambda}{\Lambda}
\bmdefine{\biPhi}{\Phi}
\bmdefine{\biSigma}{\Sigma}
\bmdefine{\biOmega}{\Omega}
\bmdefine{\biTheta}{\Theta}

\safemath{\bimA}{\biAd}
\safemath{\bimB}{\biBd}
\safemath{\bimC}{\biCd}
\safemath{\bimD}{\biDd}
\safemath{\bimE}{\biEd}
\safemath{\bimF}{\biFd}
\safemath{\bimG}{\biGd}
\safemath{\bimH}{\biHd}
\safemath{\bimI}{\biId}
\safemath{\bimJ}{\biJd}
\safemath{\bimK}{\biKd}
\safemath{\bimL}{\biLd}
\safemath{\bimM}{\biMd}
\safemath{\bimN}{\biNd}
\safemath{\bimO}{\biOd}
\safemath{\bimP}{\biPd}
\safemath{\bimQ}{\biQd}
\safemath{\bimR}{\biRd}
\safemath{\bimS}{\biSd}
\safemath{\bimT}{\biTd}
\safemath{\bimU}{\biUd}
\safemath{\bimV}{\biVd}
\safemath{\bimW}{\biWd}
\safemath{\bimX}{\biXd}
\safemath{\bimY}{\biYd}
\safemath{\bimZ}{\biZd}

\safemath{\bimDelta}{\biDelta}
\safemath{\bimLambda}{\biLambda}
\safemath{\bimPhi}{\biPhi}
\safemath{\bimSigma}{\biSigma}
\safemath{\bimOmega}{\biOmega}
\safemath{\bimTheta}{\biTheta}

\safemath{\setA}{\mathcal{A}}
\safemath{\setB}{\mathcal{B}}
\safemath{\setC}{\mathcal{C}}
\safemath{\setD}{\mathcal{D}}
\safemath{\setE}{\mathcal{E}}
\safemath{\setF}{\mathcal{F}}
\safemath{\setG}{\mathcal{G}}
\safemath{\setH}{\mathcal{H}}
\safemath{\setI}{\mathcal{I}}
\safemath{\setJ}{\mathcal{J}}
\safemath{\setK}{\mathcal{K}}
\safemath{\setL}{\mathcal{L}}
\safemath{\setM}{\mathcal{M}}
\safemath{\setN}{\mathcal{N}}
\safemath{\setO}{\mathcal{O}}
\safemath{\setP}{\mathcal{P}}
\safemath{\setQ}{\mathcal{Q}}
\safemath{\setR}{\mathcal{R}}
\safemath{\setS}{\mathcal{S}}
\safemath{\setT}{\mathcal{T}}
\safemath{\setU}{\mathcal{U}}
\safemath{\setV}{\mathcal{V}}
\safemath{\setW}{\mathcal{W}}
\safemath{\setX}{\mathcal{X}}
\safemath{\setY}{\mathcal{Y}}
\safemath{\setZ}{\mathcal{Z}}
\safemath{\emptySet}{\varnothing}

\safemath{\colA}{\mathscr{A}}
\safemath{\colB}{\mathscr{B}}
\safemath{\colC}{\mathscr{C}}
\safemath{\colD}{\mathscr{D}}
\safemath{\colE}{\mathscr{E}}
\safemath{\colF}{\mathscr{F}}
\safemath{\colG}{\mathscr{G}}
\safemath{\colH}{\mathscr{H}}
\safemath{\colI}{\mathscr{I}}
\safemath{\colJ}{\mathscr{J}}
\safemath{\colK}{\mathscr{K}}
\safemath{\colL}{\mathscr{L}}
\safemath{\colM}{\mathscr{M}}
\safemath{\colN}{\mathscr{N}}
\safemath{\colO}{\mathscr{O}}
\safemath{\colP}{\mathscr{P}}
\safemath{\colQ}{\mathscr{Q}}
\safemath{\colR}{\mathscr{R}}
\safemath{\colS}{\mathscr{S}}
\safemath{\colT}{\mathscr{T}}
\safemath{\colU}{\mathscr{U}}
\safemath{\colV}{\mathscr{V}}
\safemath{\colW}{\mathscr{W}}
\safemath{\colX}{\mathscr{X}}
\safemath{\colY}{\mathscr{Y}}
\safemath{\colZ}{\mathscr{Z}}

\safemath{\opA}{\mathbb{A}}
\safemath{\opB}{\mathbb{B}}
\safemath{\opC}{\mathbb{C}}
\safemath{\opD}{\mathbb{D}}
\safemath{\opE}{\mathbb{E}}
\safemath{\opF}{\mathbb{F}}
\safemath{\opG}{\mathbb{G}}
\safemath{\opH}{\mathbb{H}}
\safemath{\opI}{\mathbb{I}}
\safemath{\opJ}{\mathbb{J}}
\safemath{\opK}{\mathbb{K}}
\safemath{\opL}{\mathbb{L}}
\safemath{\opM}{\mathbb{M}}
\safemath{\opN}{\mathbb{N}}
\safemath{\opO}{\mathbb{O}}
\safemath{\opP}{\mathbb{P}}
\safemath{\opQ}{\mathbb{Q}}
\safemath{\opR}{\mathbb{R}}
\safemath{\opS}{\mathbb{S}}
\safemath{\opT}{\mathbb{T}}
\safemath{\opU}{\mathbb{U}}
\safemath{\opV}{\mathbb{V}}
\safemath{\opW}{\mathbb{W}}
\safemath{\opX}{\mathbb{X}}
\safemath{\opY}{\mathbb{Y}}
\safemath{\opZ}{\mathbb{Z}}
\safemath{\opZero}{\mathbb{O}}
\safemath{\identityop}{\opI}

\safemath{\veca}{\bma}
\safemath{\vecb}{\bmb}
\safemath{\vecc}{\bmc}
\safemath{\vecd}{\bmd}
\safemath{\vece}{\bme}
\safemath{\vecf}{\bmf}
\safemath{\vecg}{\bmg}
\safemath{\vech}{\bmh}
\safemath{\veci}{\bmi}
\safemath{\vecj}{\bmj}
\safemath{\veck}{\bmk}
\safemath{\vecl}{\bml}
\safemath{\vecm}{\bmm}
\safemath{\vecn}{\bmn}
\safemath{\veco}{\bmo}
\safemath{\vecp}{\bmp}
\safemath{\vecq}{\bmq}
\safemath{\vecr}{\bmr}
\safemath{\vecs}{\bms}
\safemath{\vect}{\bmt}
\safemath{\vecu}{\bmu}
\safemath{\vecv}{\bmv}
\safemath{\vecw}{\bmw}
\safemath{\vecx}{\bmx}
\safemath{\vecy}{\bmy}
\safemath{\vecz}{\bmz}

\safemath{\veczero}{\bmzero}
\safemath{\vecone}{\bmone}
\safemath{\vecxi}{\bmxi}
\safemath{\veclambda}{\bmlambda}
\safemath{\vecmu}{\bmmu}
\safemath{\vectheta}{\bmtheta}
\safemath{\vecphi}{\bmphi}
\safemath{\vecdelta}{\bmdelta}

\safemath{\matA}{\bA}
\safemath{\matB}{\bB}
\safemath{\matC}{\bC}
\safemath{\matD}{\bD}
\safemath{\matE}{\bE}
\safemath{\matF}{\bF}
\safemath{\matG}{\bG}
\safemath{\matH}{\bH}
\safemath{\matI}{\bI}
\safemath{\matJ}{\bJ}
\safemath{\matK}{\bK}
\safemath{\matL}{\bL}
\safemath{\matM}{\bM}
\safemath{\matN}{\bN}
\safemath{\matO}{\bO}
\safemath{\matP}{\bP}
\safemath{\matQ}{\bQ}
\safemath{\matR}{\bR}
\safemath{\matS}{\bS}
\safemath{\matT}{\bT}
\safemath{\matU}{\bU}
\safemath{\matV}{\bV}
\safemath{\matW}{\bW}
\safemath{\matX}{\bX}
\safemath{\matY}{\bY}
\safemath{\matZ}{\bZ}
\safemath{\matzero}{\bmzero}

\safemath{\matDelta}{\bDelta}
\safemath{\matLambda}{\bLambda}
\safemath{\matPhi}{\bPhi}
\safemath{\matSigma}{\bSigma}
\safemath{\matOmega}{\bOmega}
\safemath{\matTheta}{\bTheta}

\safemath{\matidentity}{\matI}
\safemath{\matone}{\matO}

\safemath{\rnda}{A}
\safemath{\rndb}{B}
\safemath{\rndc}{C}
\safemath{\rndd}{D}
\safemath{\rnde}{E}
\safemath{\rndf}{F}
\safemath{\rndg}{G}
\safemath{\rndh}{H}
\safemath{\rndi}{I}
\safemath{\rndj}{J}
\safemath{\rndk}{K}
\safemath{\rndl}{L}
\safemath{\rndm}{M}
\safemath{\rndn}{N}
\safemath{\rndo}{O}
\safemath{\rndp}{P}
\safemath{\rndq}{Q}
\safemath{\rndr}{R}
\safemath{\rnds}{S}
\safemath{\rndt}{T}
\safemath{\rndu}{U}
\safemath{\rndv}{V}
\safemath{\rndw}{W}
\safemath{\rndx}{X}
\safemath{\rndy}{Y}
\safemath{\rndz}{Z}

\safemath{\rveca}{\bimA}
\safemath{\rvecb}{\bimB}
\safemath{\rvecc}{\bimC}
\safemath{\rvecd}{\bimD}
\safemath{\rvece}{\bimE}
\safemath{\rvecf}{\bimF}
\safemath{\rvecg}{\bimG}
\safemath{\rvech}{\bimH}
\safemath{\rveci}{\bimI}
\safemath{\rvecj}{\bimJ}
\safemath{\rveck}{\bimK}
\safemath{\rvecl}{\bimL}
\safemath{\rvecm}{\bimM}
\safemath{\rvecn}{\bimN}
\safemath{\rveco}{\bomO}
\safemath{\rvecp}{\bimP}
\safemath{\rvecq}{\bimQ}
\safemath{\rvecr}{\bimR}
\safemath{\rvecs}{\bimS}
\safemath{\rvect}{\bimT}
\safemath{\rvecu}{\bimU}
\safemath{\rvecv}{\bimV}
\safemath{\rvecw}{\bimW}
\safemath{\rvecx}{\bimX}
\safemath{\rvecy}{\bimY}
\safemath{\rvecz}{\bimZ}

\safemath{\rvecxi}{\bmxi}
\safemath{\rveclambda}{\bmlambda}
\safemath{\rvecmu}{\bmmu}
\safemath{\rvectheta}{\bmtheta}
\safemath{\rvecphi}{\bmphi}

\safemath{\rmatA}{\bimA}
\safemath{\rmatB}{\bimB}
\safemath{\rmatC}{\bimC}
\safemath{\rmatD}{\bimD}
\safemath{\rmatE}{\bimE}
\safemath{\rmatF}{\bimF}
\safemath{\rmatG}{\bimG}
\safemath{\rmatH}{\bimH}
\safemath{\rmatI}{\bimI}
\safemath{\rmatJ}{\bimJ}
\safemath{\rmatK}{\bimK}
\safemath{\rmatL}{\bimL}
\safemath{\rmatM}{\bimM}
\safemath{\rmatN}{\bimN}
\safemath{\rmatO}{\bimO}
\safemath{\rmatP}{\bimP}
\safemath{\rmatQ}{\bimQ}
\safemath{\rmatR}{\bimR}
\safemath{\rmatS}{\bimS}
\safemath{\rmatT}{\bimT}
\safemath{\rmatU}{\bimU}
\safemath{\rmatV}{\bimV}
\safemath{\rmatW}{\bimW}
\safemath{\rmatX}{\bimX}
\safemath{\rmatY}{\bimY}
\safemath{\rmatZ}{\bimZ}

\safemath{\rmatDelta}{\bimDelta}
\safemath{\rmatLambda}{\bimLambda}
\safemath{\rmatPhi}{\bimPhi}
\safemath{\rmatSigma}{\bimSigma}
\safemath{\rmatOmega}{\bimOmega}
\safemath{\rmatTheta}{\bimTheta}

\usepackage{amssymb}
\usepackage{amsfonts}
\usepackage{mathrsfs}
\usepackage{xspace}
\usepackage{bm}
\usepackage{fancyref}
\usepackage{textcomp}

\usepackage{multirow}
\usepackage{stmaryrd}

\newenvironment{textbmatrix}{	\setlength{\arraycolsep}{2.5pt}%
								\left[\begin{matrix}}{\end{matrix}\right]%
								\raisebox{0.08ex}{\vphantom{M}}}

\def\be{\begin{equation}}
\def\ee{\end{equation}}
\def\een{\nonumber \end{equation}}
\def\mat{\begin{bmatrix}}
\def\emat{\end{bmatrix}}
\def\btm{\begin{textbmatrix}}
\def\etm{\end{textbmatrix}}

\def\ba#1\ea{\begin{align}#1\end{align}}
\def\bas#1\eas{\begin{align*}#1\end{align*}}
\def\bs#1\es{\begin{split}#1\end{split}}
\def\bg#1\eg{\begin{gather}#1\end{gather}}
\def\bml#1\eml{\begin{multline}#1\end{multline}}
\def\bi#1\ei{\begin{itemize}#1\end{itemize}}

\safemath{\dirac}{\delta}					% Dirac delta
\safemath{\krond}{\dirac}					% Kronecker delta
\safemath{\upto}{\uparrow}
\safemath{\downto}{\downarrow}
\safemath{\iu}{j}							% imaginary unit
\safemath{\ev}{\lambda}						% eigenvalue
\safemath{\hilseqspace}{l^{2}}				% Hilbert sequence space
\newcommand{\banachfunspace}[1]{\setL^{#1}}	% Banach function space
\safemath{\hilfunspace}{\banachfunspace{2}}	% Hilbert function space
\safemath{\SNR}{\textit{SNR}} 				% signal to noise ratio
\safemath{\PAR}{\textit{PAR}} 				% signal to noise ratio
\safemath{\No}{N_0}							% noise spectral density
\safemath{\Es}{E_s}							% energy per symbol
\safemath{\Eb}{E_b}							% energy per bit
\safemath{\EbNo}{\frac{\Eb}{\No}}
\safemath{\EsNo}{\frac{\Es}{\No}}

\DeclareMathOperator{\CHop}{\ensuremath{\opH}} % channel operator
\safemath{\tvir}{\rndh_{\CHop}}				% time-varying impulse response
\safemath{\tvtf}{\rndl_{\CHop}}				% 	-''- transfer function
\safemath{\spf}{\rnds_{\CHop}}				% spreading function
\safemath{\bff}{H_{\CHop}}					% bi-freuqency function

\safemath{\ircf}{r_{h}}						% impulse response correlation fn.
\safemath{\tftvcf}{r_{s}}					% scattering function
\safemath{\tfcf}{r_{l}}						% time-frequency correlation fn.
\safemath{\bfcf}{r_{H}}						% bi-frequency correlation fn.

\safemath{\tcorr}{c_h}						% time-correlation function
\safemath{\scf}{c_{s}}						% spreading function
\safemath{\tfcorr}{c_{l}}					% transfer-function correlation
\safemath{\fcorr}{c_{H}}						% frequency-correlation function

\safemath{\mi}{I}							% mutual information
\safemath{\capacity}{C}						% capacity

\safemath{\normal}{\mathcal{N}}			% normal distribution
\safemath{\jpg}{\mathcal{CN}}			% jointly proper Gaussian
\safemath{\mchain}{\leftrightarrow}		% Markov chain
\safemath{\dB}{\,\mathrm{dB}}
\safemath{\dBm}{\,\mathrm{dBm}}
\safemath{\Hz}{\,\mathrm{Hz}}
\safemath{\kHz}{\,\mathrm{kHz}}
\safemath{\MHz}{\,\mathrm{MHz}}
\safemath{\GHz}{\,\mathrm{GHz}}
\safemath{\s}{\,\mathrm{s}}
\safemath{\ms}{\,\mathrm{ms}}
\safemath{\mus}{\,\mathrm{\text{\textmu}s}}
\safemath{\ns}{\,\mathrm{ns}}
\safemath{\ps}{\,\mathrm{ps}}
\safemath{\meter}{\,\mathrm{m}}
\safemath{\mm}{\,\mathrm{mm}}
\safemath{\cm}{\,\mathrm{cm}}
\safemath{\m}{\,\mathrm{m}}
\safemath{\W}{\,\mathrm{W}}
\safemath{\mW}{\, \mathrm{mW}}
\safemath{\J}{\,\mathrm{J}}
\safemath{\K}{\,\mathrm{K}}
\safemath{\bit}{\,\mathrm{bit}}
\safemath{\nat}{\,\mathrm{nat}}

\safemath{\define}{\triangleq}			% definition

\safemath{\equivalent}{\sim}
\safemath{\distas}{\sim}					% distributed according to
\safemath{\sdiff}{\Delta}				% symmetric set difference

\safemath{\reals}{\mathbb{R}}
\safemath{\positivereals}{\reals_{+}}
\safemath{\integers}{\mathbb{Z}}
\safemath{\posint}{\integers_{+}}
\safemath{\naturals}{\mathbb{N}}
\safemath{\posnaturals}{\naturals_{+}}
\safemath{\complexset}{\mathbb{C}}
\safemath{\rationals}{\mathbb{Q}}

\newcommand*{\fancyrefapplabelprefix}{app}		% Appendix
\newcommand*{\fancyrefthmlabelprefix}{thm}		% Theorem
\newcommand*{\fancyreflemlabelprefix}{lem}		% Lemma
\newcommand*{\fancyrefcorlabelprefix}{cor}		% Corollary
\newcommand*{\fancyrefdeflabelprefix}{def}		% Definition
\newcommand*{\fancyrefproplabelprefix}{prop}		% Proposition
\newcommand*{\fancyrefexmpllabelprefix}{exmpl}
\newcommand*{\fancyrefalglabelprefix}{alg}		% Algorithm
\newcommand*{\fancyreftbllabelprefix}{tbl}		% Algorithm

\frefformat{vario}{\fancyrefseclabelprefix}{Sec.~#1}
\frefformat{vario}{\fancyrefthmlabelprefix}{Thm.~#1}
\frefformat{vario}{\fancyreftbllabelprefix}{Tbl.~#1}
\frefformat{vario}{\fancyreflemlabelprefix}{Lem.~#1}
\frefformat{vario}{\fancyrefcorlabelprefix}{Corr.~#1}
\frefformat{vario}{\fancyrefdeflabelprefix}{Def.~#1}
\frefformat{vario}{\fancyreffiglabelprefix}{Fig.~#1}
\frefformat{vario}{\fancyrefapplabelprefix}{App.~#1}
\frefformat{vario}{\fancyrefeqlabelprefix}{(#1)}
\frefformat{vario}{\fancyrefproplabelprefix}{Prop.~#1}
\frefformat{vario}{\fancyrefexmpllabelprefix}{Ex.~#1}
\frefformat{vario}{\fancyrefalglabelprefix}{Alg.~#1}

\newcommand{\WiFi}{\mbox{Wi-Fi}\xspace}

\safemath{\dictab}{[\,\dicta\,\,\dictb\,]}

\safemath{\ysig}{\bmy}
\safemath{\ysighat}{\hat{\ysig}}
\safemath{\ysigdim}{M}
\safemath{\xsig}{\bmx}
\safemath{\xsigdim}{N}
\safemath{\nx}{n_x}
\safemath{\zsig}{\bmz}
\safemath{\zsigdim}{\ysigdim}
\safemath{\rsig}{\bmr}
\safemath{\Adict}{\bA}
\safemath{\Adicttilde}{\widetilde{\Adict}}
\safemath{\Adictdim}{\outputdim\times\xsigdim}
\safemath{\avec}{\bma}
\safemath{\avectilde}{\tilde{\avec}}
\safemath{\Bdict}{\bB}
\safemath{\Bdicttilde}{\widetilde{\Bdict}}
\safemath{\Cdict}{\bC}
\safemath{\cvec}{\bmc}
\safemath{\Ddict}{\bD}
\safemath{\Ddictdim}{\ysigdim\times\xsigdim}
\safemath{\dvec}{\bmd}
\safemath{\Ddicttilde}{\widetilde{\bD}}
\safemath{\Bonb}{\bB}
\safemath{\bvec}{\bmb}
\safemath{\Bonbdim}{\ysigdim\times\ysigdim}
\safemath{\noise}{\bmn}
\safemath{\noisedim}{\ysigim}
\safemath{\err}{\bme}
\safemath{\errdim}{\ysigdim}
\safemath{\errset}{\setE}
\safemath{\nerr}{n_e}
\safemath{\delop}{\bP_\errset}
\safemath{\delopc}{\bP_{{\errset}^c}}

\safemath{\cplxi}{\imath}
\safemath{\cplxj}{\jmath}
\safemath{\dict}{\matD}
\safemath{\inputdim}{N}		% number of columns of dictionary D
\safemath{\outputdim}{M}		%number of rows of dictionary D
\safemath{\sparsity}{S}	%sparsity
\safemath{\inputdimA}{{N_a}}	%total number of elements in dictionary A
\safemath{\inputdimB}{{N_b}}	%total number of elements in dictionary B
\safemath{\elemA}{{n_a}}	%number of elements chosen from dictionary A
\safemath{\elemB}{{n_b}}	%number of elements chosen from dictionary B
\safemath{\resA}{\matR_a}	%restriction map to elements of dictionary A
\safemath{\resB}{\matR_b}	%restriction map to elements of dictionary B
\safemath{\subD}{\matS} %subdictionary
\safemath{\subA}{\matS_a} %subdictionary part of A
\safemath{\subB}{\matS_b} %subdictionary part of B
\safemath{\dicta}{\matA} 	% first subdictionary
\safemath{\dictb}{\matB} 	% second subdictionary
\safemath{\hollowS}{H}
\safemath{\hollowA}{H_a}
\safemath{\hollowB}{H_b}
\safemath{\cross}{Z}
\safemath{\coh}{\mu_d}			% coherence dictionary
\safemath{\coha}{\mu_a}			% coherence first subdictionary
\safemath{\cohb}{\mu_b}			% coherence second subdictionary
\safemath{\mubs}{\nu}	%block sub-coherence
\safemath{\cohm}{\mu_m} %mutual coherence
\safemath{\dictset}{\setD}	% set of dictionaries
\safemath{\dictsetp}{\dictset(\coh,\coha,\cohb)}	% set of dictionaries parametrized
\safemath{\dictsetgen}{\dictset_\text{gen}}
\safemath{\dictsetgenp}{\dictsetgen(\coh)}
\safemath{\dictsetonb}{\dictset_\text{onb}}
\safemath{\dictsetonbp}{\dictsetonb(\coh)}

\safemath{\leftside}{U}
\safemath{\rightsideA}{R_a}
\safemath{\rightsideB}{R_b}

\safemath{\indexS}{\setI_S} %set of indices participating in sub-dictionary S

\safemath{\na}{n_a}			% cardinality of set of linearly independent columns of first dictionary
\safemath{\nb}{n_b}			% cardinality of set of linearly independent columns of second dictionary
\safemath{\coeffa}{p_i}	%coefficients for columns of A
\safemath{\coeffb}{q_j}	%coefficients for columns of B
\safemath{\seta}{\setP}		% set of linearly independent columns of A
\safemath{\setb}{\setQ}     % set of linearly independent columns of B
\safemath{\setw}{\setW}	%set of n largest elements of w
\safemath{\setz}{\setZ}	%set of L-n largest elements of z
\safemath{\cola}{\veca}		% generic element of the dictionary A
\safemath{\colb}{\vecb}		% generic element of the dictionary B
\safemath{\cold}{\vecd}		% generic element of the dictionary D
\safemath{\inputvec}{\vecx} 	%coefficient vector (input)
\safemath{\error}{\vece}	%error vector
\safemath{\noiseout}{\vecz} 	%noisy output vector
\safemath{\inputvecel}{x}
\safemath{\inputveca}{\vecx_a}
\safemath{\inputvecb}{\vecx_b}
\safemath{\outputvec}{\vecy}	%output of Dictionary
\safemath{\lambdamin}{\lambda_{\mathrm{min}}}
\safemath{\elltwo}{\ell_2}
\safemath{\ellone}{\ell_1}
\safemath{\ellzero}{\ell_0}
\safemath{\ellinf}{\ell_\infty}
\safemath{\ellinftilde}{\ell_{\widetilde\infty}}
\safemath{\licard}{Z(\coh,\coha,\cohb)}
\safemath{\xsol}{\hat{x}}
\safemath{\xbord}{x_b}		%Solution at the border
\safemath{\xstat}{x_s}		%Solution stationary in l0 prob
\safemath{\xstatLone}{\tilde{x}_s}
\safemath{\order}{\mathcal{O}} %order notation (big O)
\safemath{\scales}{\Theta} %scales as
\safemath{\ones}{\mathbf{1}} %all ones matrix
\safemath{\zeroes}{\mathbf{0}} %all zeroes matrix
\safemath{\thlone}{\kappa(\coh,\cohb)} %treshold l1 problem
\safemath{\constoneA}{\delta} %constant in l1 theorem to save space
\safemath{\constoneB}{\epsilon} %constant in l1 theorem to save space
\safemath{\nlarge}{L}				   %num large elements
\safemath{\sumlarge}{S_\nlarge}
\safemath{\maxlarger}{P_\nlarge}	   % maximum in Gribonval and Nielsen
\safemath{\Pzero}{\textrm{P0}}	
\safemath{\Pone}{\textrm{P1}}
\safemath{\vecfir}{\vecw}			 % \vecv element of the kernel of the dictionary, \vecv=[\vecfir \vecsec]
\safemath{\vecsec}{\vecz}
\safemath{\elvecfir}{w}              % element of vecfir
\safemath{\elvecsec}{z}				 % element of vecsec
\safemath{\nlargefir}{n}
\safemath{\normout}{\gamma}
\safemath{\auxfun}{h}
\safemath{\supp}{\textrm{supp}}%support

\safemath{\indexa}{\ell}
\safemath{\indexb}{r}
\safemath{\indexc}{i}
\safemath{\indexd}{j}

\safemath{\project}{P}%projector

\safemath{\firstslotset}{\setU_1}  % set of UEs for first slot
\safemath{\secondslotset}{\setU_2} % set of UEs for second slot
\safemath{\randomset}{\setS} % generic random scheduling

\safemath{\Tran}{\textnormal{T}}
\safemath{\Herm}{\textnormal{H}}

\newcommand*{\fancyreflstlabelprefix}{lst}
\fancyrefaddcaptions{english}{%
  \providecommand*{\freflstname}{Listing}%
}
\frefformat{vario}{\fancyreflstlabelprefix}{%
  \freflstname\fancyrefdefaultspacing#1#3%
}

\begin{document}

\title{CSI-Based User Positioning, Channel Charting, and Device Classification with an NVIDIA 5G Testbed}
\author{

    \IEEEauthorblockN{Reinhard Wiesmayr$^\text{1}$, Frederik Zumegen$^\text{1}$, Sueda Taner$^\text{1}$, Chris Dick$^\text{2}$, and Christoph Studer$^\text{1}$}\\[0.2cm]
    \textit{$^\textnormal{1}$ETH Zurich, $^\textnormal{2}$NVIDIA; e-mail: wiesmayr@iis.ee.ethz.ch}\\[0.1cm]
    \thanks{The authors thank Gian Marti for comments and suggestions. The authors also thank Ariane Frommenwiler and Georg Rutishauser for their help during CSI collection.}
    \thanks{We acknowledge NVIDIA for their sponsorship of this research.}
    \thanks{This work was supported in part by the Swiss National Science Foundation (SNSF) grant 200021\_207314 and by CHIST-ERA grant for the project CHASER (CHIST-ERA-22-WAI-01) through the SNSF grant 20CH21\_218704.}
    }

\maketitle

\glsresetall

\begin{abstract}

\Gls{CSI}-based sensing will play a key role in future cellular systems. 
However, no \gls{CSI} dataset has been published from a real-world 5G NR system that facilitates the development and validation of suitable sensing algorithms.
To close this gap, we publish three real-world wideband multi-antenna multi-\gls{ORU} \gls{CSI} datasets from the 5G NR uplink channel: an indoor lab/office room dataset, an outdoor campus courtyard dataset, and a device classification dataset with six \gls{COTS} \glspl{UE}.
These datasets have been recorded using a software-defined 5G NR testbed based on NVIDIA \gls{ARC-OTA} with \gls{COTS} hardware, which we have deployed at ETH Zurich.
We demonstrate the utility of these datasets for three \gls{CSI}-based sensing tasks: neural \gls{UE} positioning, channel charting in real-world coordinates, and closed-set device classification.
For all these tasks, our results show high accuracy: neural \gls{UE} positioning achieves 0.6\,cm (indoor) and 5.7\,cm (outdoor) mean absolute error, channel charting in real-world coordinates achieves 73\,cm mean absolute error (outdoor), and device classification achieves 99\,\% (same day) and 95\,\% (next day) accuracy.
The \gls{CSI} datasets, ground-truth \gls{UE} position labels, \gls{CSI} features, and simulation code are publicly available at \url{https://caez.ethz.ch}

\end{abstract}

\glsresetall

\section{Introduction}

Cellular systems continuously measure \gls{CSI} for coherent data detection and beamforming. 
The \gls{CSI} acquired at \glspl{BS} can be reused for sensing tasks, such as off-device neural positioning \cite{pulkkinen2011semi,ferrand2020dnn,gonultacs2021csi} and device classification \cite{sharaf2016authentication, shen2022towards}, which are applications that are expected to play a major role in next-generation wireless systems~\cite{6g_localization_sensing}.
Since the development and evaluation of such \gls{CSI}-based sensing methods with purely synthetic data (e.g., from a ray-tracer) can lead to overly optimistic results \cite{bock2025wireless}, a performance validation with real-world \gls{CSI} measurements is essential. 
In particular, algorithms for sixth-generation (6G) wireless systems should be developed and validated on 5G New Radio (NR) systems, since 5G NR is the latest communication standard most similar to 6G.\footnote{For example, 6G will implement similar \gls{OFDM} waveforms as in 5G NR~\cite{3gpp_RP25nnnn}.}
To the best of our knowledge, however, no publicly available 5G NR \gls{CSI} dataset from a 
commercial 5G \gls{BS} currently exists.
The literature reports only a few real-world experiments for next-generation positioning or device classification systems, such as \cite{pileggi20235g} and \cite{fu2023radio}, respectively.
Instead, prior work utilizes \gls{CSI} datasets from channel sounders~\cite{dichasus2021} or custom\footnote{Custom testbeds are specific solutions for the purpose of \gls{CSI} collection and often do not implement the full protocol stack. In contrast, commercial communication systems are primarily built to facilitate \gls{UE} communication, implement the complete protocol stack, and deliver \gls{CSI} measurements only as a side-product from real network traffic.} \WiFi testbeds~\cite{euchner2024espargos,zumegen2024software} that are less representative for 6G.
For 5G/6G research, we are only aware of one dataset of \gls{DL} \gls{CSI} published in~\cite{FraunhoferIIS_FingerprintingDataset2025}, which was obtained from a custom 5G testbed built with software-defined radios.

\subsection{Contributions}
\begin{figure}[t]
    \centering
    \includegraphics[width=.6\columnwidth]{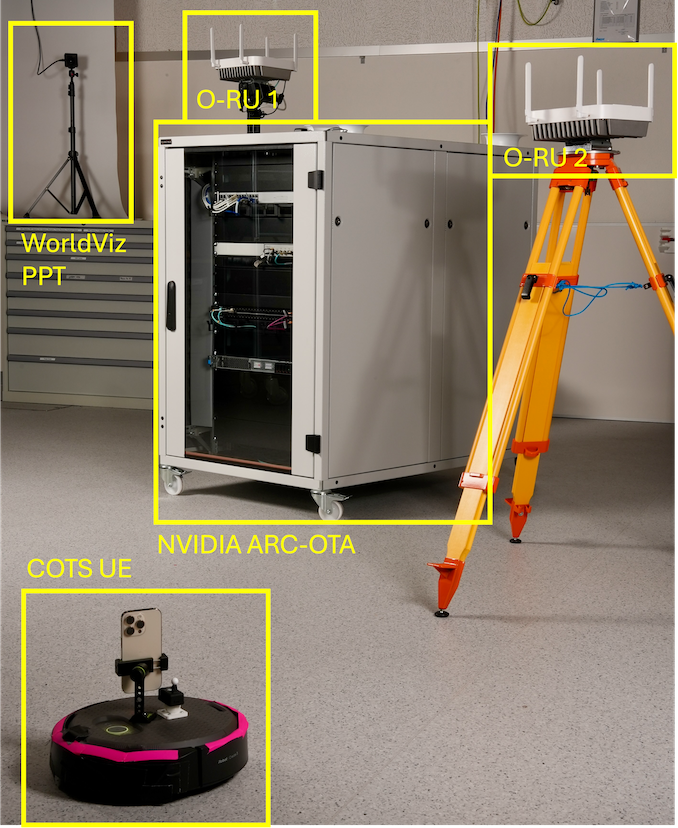}
\caption{Photo of the ETH Zurich 5G NR testbed with \gls{COTS} \gls{UE} and \gls{COTS} \glspl{ORU}. The 5G network stack runs in real-time on the NVIDIA \gls{ARC-OTA} system~\cite{nvidia_ARC-OTA_2025}. The WorldViz PPT~\cite{worldviz} is used for \gls{UE} position tracking.}
    \label{fig:mobile_testbed}
\end{figure}

We collect \gls{UL} \gls{CSI} from real-world 5G NR network traffic on a standard-compliant 5G NR system with \gls{COTS} \glspl{UE} and four multi-antenna \gls{COTS} \glspl{ORU} with four antennas each.
The system is depicted in \fref{fig:mobile_testbed} and builds upon the recently introduced NVIDIA \gls{ARC-OTA}~\cite{nvidia_ARC-OTA_2025} platform. This is a software-defined full-stack 5G NR system that enables data collection and \gls{CSI} extraction from real-time 5G NR traffic.
With this system, we collected three different datasets, collectively referred to as \gls{CAEZ}: (i) an indoor lab and office room dataset, (ii) an outdoor campus courtyard dataset, and (iii) a device classification dataset with \gls{CSI} from six \gls{COTS} \glspl{UE}.
To demonstrate the utility of these datasets,  we showcase three different \gls{CSI}-based sensing tasks: (i) neural \gls{UE} positioning, (ii) channel charting, and (iii) closed-set device classification. 
For all three sensing tasks, we achieve excellent accuracy.

\subsection{Relevant Prior Art}

Existing alternatives for measuring real-world \gls{CSI} include channel sounders as well as custom \WiFi and 5G NR testbeds.
Channel sounders, such as DICHASUS \cite{dichasus2021}, employ custom protocols with specialized hardware.
\WiFi testbeds, including ESPARGOS \cite{euchner2024espargos} and the system in \cite{zumegen2024software}, often build on inexpensive \gls{COTS} \glspl{UE} and operate in unlicensed \gls{ISM} bands where potential interference can affect measurements.
Furthermore, \WiFi utilizes different waveforms (i.e., different \gls{OFDM} numerology and framing) and protocols than 5G and emerging 6G systems, and it applies random access mechanisms without the centralized schedulers present in 5G systems.
The custom 5G testbeds presented in~\cite{FraunhoferIIS_FingerprintingDataset2025,pileggi20235g} collect \gls{DL} \gls{CSI} from multiple custom 5G NR \glspl{BS} and a custom \gls{UE} leveraging software-defined radios.
In contrast, we collect \gls{UL} \gls{CSI} from real uplink network traffic using a fully 5G NR-compliant system in licensed spectrum with multiple \gls{COTS} \glspl{ORU} and \gls{COTS} \glspl{UE}.

\gls{CSI}-based \gls{UE} positioning with \glspl{NN}, called neural positioning, is a supervised positioning method that has been validated with \WiFi datasets~\cite{gonultacs2021csi, zumegen2024software} and large-scale 4G outdoor measurements \cite{ferrand2020dnn}.
However, most of the \WiFi and 4G measurements are not publicly available, except for some small-scale indoor experiments with sparse spatial coverage~\cite{gao2020crisloc,gassner2021opencsi,FraunhoferIIS_FingerprintingDataset2025}.
We close this gap by publishing 5G NR \gls{CSI} datasets with dense spatial coverage and validate their utility for neural positioning using the method from~\cite{gonultacs2021csi}.

Channel charting is an emerging self-supervised pseudo-positioning method that has been demonstrated on 
public \WiFi datasets \cite{euchner2022improving, euchner2023augmenting, euchner2024espargos} and on a large-scale 4G massive \gls{MIMO} outdoor dataset \cite{ferrand2021triplet} with 64 antennas and 10\,MHz bandwidth.
However, the 4G dataset is not publicly available.
We close this gap by applying triplet-loss-based channel charting \cite{ferrand2021triplet} and channel charting in real-world coordinates \cite{taner2025channel} on distributed \gls{MIMO} indoor and outdoor 5G \gls{CSI} datasets with four multi-antenna \glspl{ORU} operating at 100\,MHz bandwidth.
Furthermore, we publish the simulation code along with the \gls{CAEZ} datasets. 

\begin{figure}[t]
    \centering
    \includegraphics[width=.95\columnwidth]{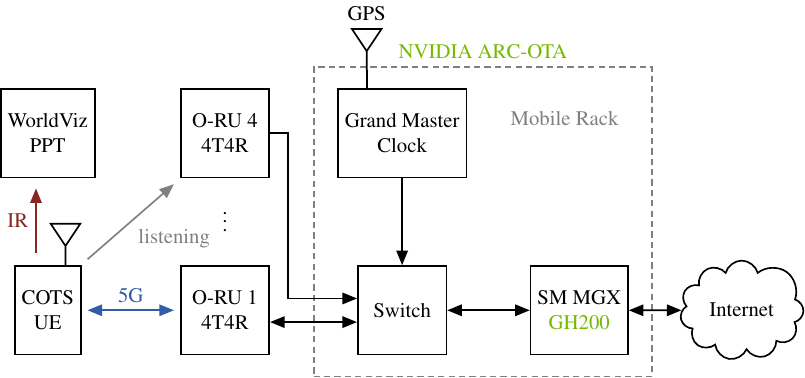}
    \caption{System diagram of the ETH Zurich 5G NR testbed.}  
    \label{fig:system_overview}
\end{figure}

Device classification has been demonstrated on LoRa WAN~\cite{shen2021radio, shen2022towards}, \WiFi \cite{mazokha2025mobrffi}, and the 5G control channel \cite{fu2023radio}.
However, LoRa WAN and \WiFi utilize different waveforms and protocols than 5G and emerging 6G systems.
Furthermore, the work in \cite{fu2023radio} captures 5G control-channel waveforms only with static \gls{UE} positions.
We close this gap by collecting \gls{CSI} from the 5G NR \gls{PUSCH} with \glspl{UE} at randomly varying locations.
As a result, we enable location-insensitive device classification with \gls{RFFI} features inspired by recent work in \gls{CSI} obfuscation \cite{stephan2025csi}.
We apply the closed-set classification pipeline published together with the simulation code of the open-set classification method presented~in~\cite{shen2022towards}.

\section{ETH Zurich 5G NR Testbed}

We developed a 5G NR testbed at ETH Zurich to collect real-world \gls{CSI} measurements in licensed spectrum, specifically the full Swiss private 5G band, i.e., 100\,MHz of the 5G NR N78 band centered at 3.45\,GHz.
The testbed builds upon NVIDIA \gls{ARC-OTA} \cite{nvidia_ARC-OTA_2025} and is a full-stack software-defined 5G system with \gls{COTS} \glspl{UE} and four \gls{COTS} \glspl{ORU}, where one \gls{ORU} is used for 5G communication, whereas the other three operate as passive listeners
(a system diagram is shown in \fref{fig:system_overview}).
All components (except for the UEs) are connected via a fiber optical switch.
A Supermicro NVIDIA MGX GH200 server runs the full-stack 5G system, comprising the NVIDIA Aerial L1, \gls{OAI} L2, and \gls{OAI} core network.
A \gls{PTP} grand master clock with \gls{GNSS} time reference synchronizes the fiber-optical network, i.e., the system clock of the \glspl{ORU} and the GH200 server.
The GH200 server also runs NVIDIA DataLake to collect \gls{FH} I/Q samples and L2 protocol data (FAPI) of received \gls{PUSCH} slots from all four \glspl{ORU} and all connected \glspl{UE}, respectively.
The GH200 server obtains ground-truth \gls{UE} positions from an external WorldViz precision position tracking (PPT) system~\cite{worldviz},
which localizes the UEs using six infrared cameras that track infrared markers mounted on the \gls{UE}-carrying vehicle.
\fref{tbl:system_parameters} lists the most important system parameters. 
Our system configuration ensures that we measure \gls{CSI} samples at least every 10\,ms or 20\,ms, depending on the scenario, and at most every 2.5\,ms; this is a consequence of the selected \gls{TDD} pattern, subcarrier spacing, and L2 scheduler configuration.

\begin{table}[t]
\centering
\caption{Key system parameters of the ETH Zurich 5G NR testbed.}
\renewcommand{\arraystretch}{1.1}
\begin{tabular}{@{}lc@{}}
\toprule
\textbf{Parameter} & \textbf{Value}\\
\midrule
Communication standard & 3GPP Rel. 15 \\
Carrier frequency & 3.45\,GHz \\
Bandwidth & 100\,MHz \\
Active subcarriers $W$ & 3'276 \\
Subcarrier spacing & 30\,kHz \\
\# of \glspl{ORU} $O$ & 4 \\
Antennas per \gls{ORU} $B$ & 4 (4T4R) \\
Configured \gls{ORU} Tx. power & 1\,W \\
Target SNR in \gls{PUSCH} & 28\,dB \\
\gls{TDD} pattern & 3DSU \\
\bottomrule
\end{tabular}
\label{tbl:system_parameters}
\end{table}

\section{CAEZ: CSI Acquisition at ETH Zurich}

We now describe the three 5G \gls{CSI} datasets acquired at ETH Zurich, collectively referred to as \gls{CAEZ}-5G.
A summary table with the key information is provided in \fref{tbl:datasets}.
The \gls{CAEZ} datasets and \gls{CSI} features are available at \url{https://caez.ethz.ch}.

\begin{table}[t]
\caption{Summary of measured \gls{CAEZ} datasets.}
\label{tbl:datasets}
\centering
\renewcommand{\arraystretch}{1.1}
\resizebox{\columnwidth}{!}{
\begin{tabular}{@{}lccc@{}}
\toprule
\textbf{\gls{CAEZ}-5G} & \textbf{INDOOR} & \textbf{OUTDOOR} & \textbf{DEV-CLASS} \\
\midrule
Duration & 1\,h\,47\,min & 1\,h\,38\,min & 6\,$\times$\,(2\,\text{min}\,+\,30\,\text{s}) \\
Area & 3.5\,m\,$\times$\,3.5\,m & 10\,m\,$\times$\,10\,m & 4\,m\,$\times$\,4\,m \\
\# of samples & 338'981 & 303'189 & 83'619\,+\,21'805 \\
\gls{UE} type & Quectel RMU500EK & \gls{UE} 4 & \gls{UE} 1a--5\\ 
Vehicle & vacuum robot & custom robot & rot. table\,+\,human \\
Position tagged & yes & yes & no \\
PUSCH every & 20\,ms & 20\,ms & 10\,ms \\
\bottomrule
\end{tabular}
}
\end{table}

\begin{figure}[t]
  \centering
  \subfigure[]{
    \includegraphics[height=3.2cm]{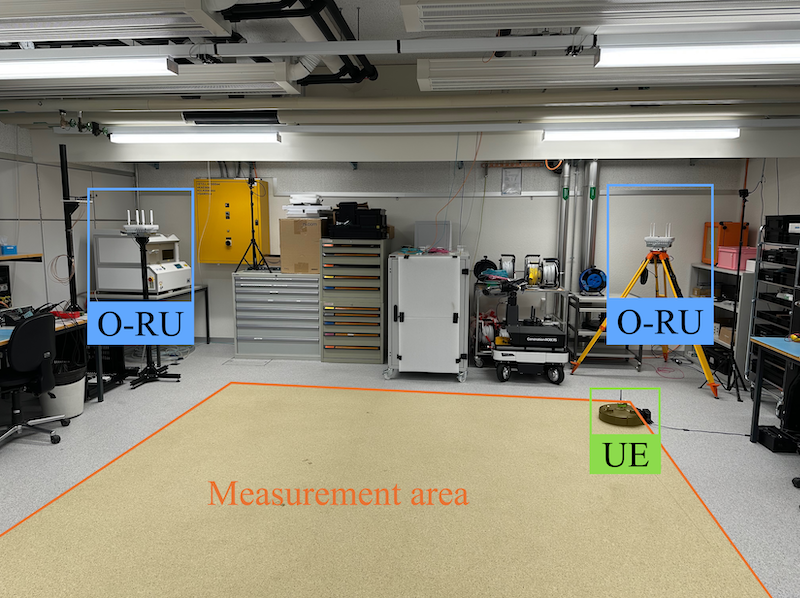}\label{fig:indoor_overview}%
  }\hfill
  \subfigure[]{
    \includegraphics[height=3.2cm]{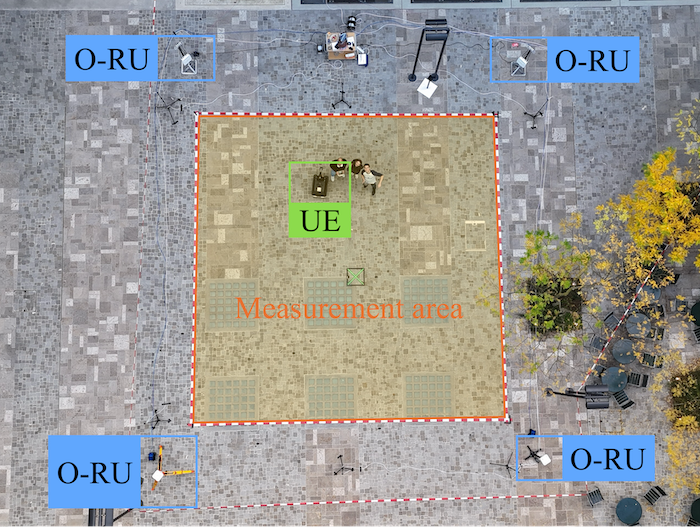}\label{fig:outdoor_overview}%
  }
  \subfigure[]{
    \includegraphics[height=3.25cm]{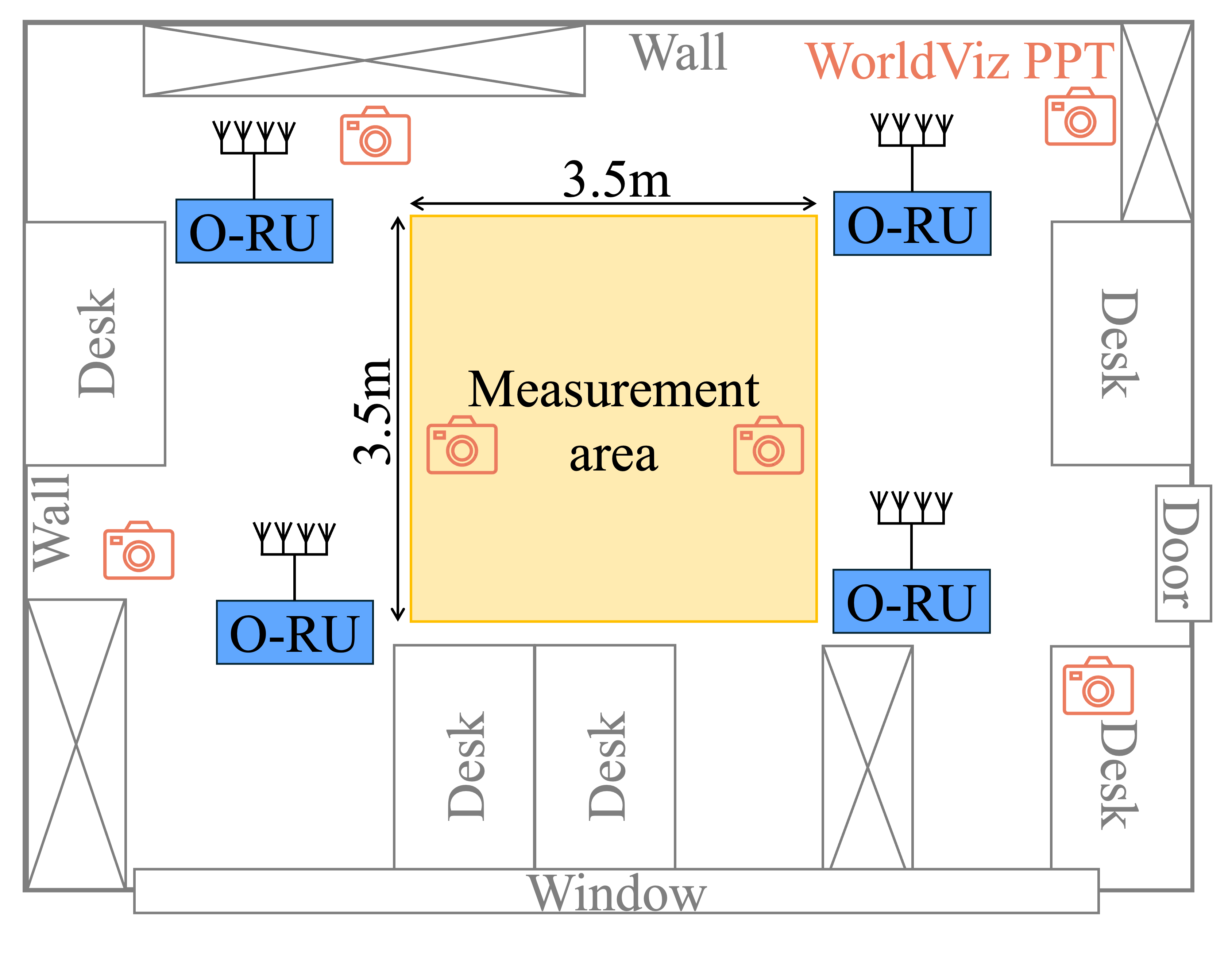}\label{fig:indoor_plan}%
  }\hfill
  \subfigure[]{
    \includegraphics[height=3.25cm]{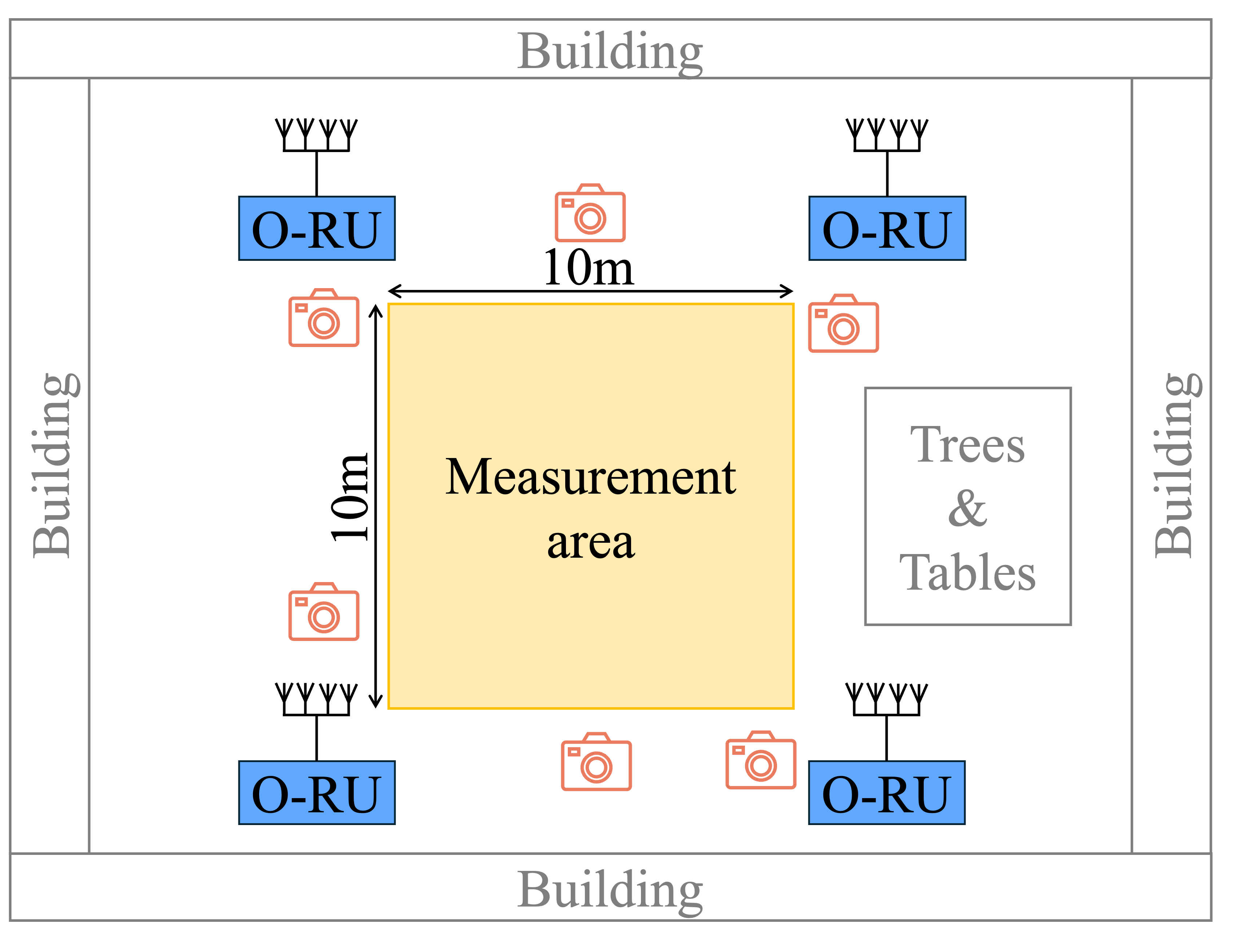}\label{fig:outdoor_plan}%
  }
  \caption{Measurement setups (top row) and floor plans (bottom row) of the \gls{CAEZ}-5G-INDOOR (left) and \gls{CAEZ}-5G-OUTDOOR (right) datasets.}
  \label{fig:caez_5g_indoor_outdoor}
\end{figure}

\subsection{CAEZ-5G-INDOOR}

\subsubsection{Overview}

The indoor measurement setup, acquired in a joint lab/office, is shown in \fref{fig:indoor_overview}.
As depicted in \fref{fig:indoor_plan}, the measurement area is a 3.5\,m\,$\times$\,3.5\,m squared area between the lab desks.
The \glspl{ORU} were placed at the corners of the measurement area.
Four WorldViz PPT cameras were placed around the measurement area and two cameras were placed above the measurement area.  

\subsubsection{Measurement Details}

We use an iRobot Create 3 robot platform \cite{iRobot_Create3} with a Raspberry Pi and a Quectel modem inside the robot's cargo bay.
A single antenna was connected to the ``PRX'' port of the Quectel modem; the other ports were terminated.
The antenna was mounted on top of the robot along the center axis.
The robot was controlled using random waypoint navigation, as described in \cite{muller2025neuralpos}.\footnote{In~\cite{muller2025neuralpos}, the CAEZ-5G-INDOOR dataset is called the ``5G Office'' dataset.}
Four WorldViz PPT markers were mounted on the robot to enable tracking of position and rotation. One measurement operator was present in the lab/office during \gls{CSI} collection and sometimes even walked through the measurement~area.

\subsection{CAEZ-5G-OUTDOOR}

\subsubsection{Overview}

The outdoor measurement setup is shown in \fref{fig:outdoor_overview}.
As depicted in \fref{fig:outdoor_plan}, the measurement area is a 10\,m\,$\times$\,10\,m squared area in the ETH Zurich electrical engineering campus courtyard, surrounded by multiple buildings, trees, and other obstacles.
The \glspl{ORU} were placed at the corners of the measurement area and six WorldViz PPT cameras were placed around the measurement area.

\subsubsection{Measurement Details}

\begin{figure}[t]
  \centering
  \subfigure[]{
    \includegraphics[height=3.3cm]{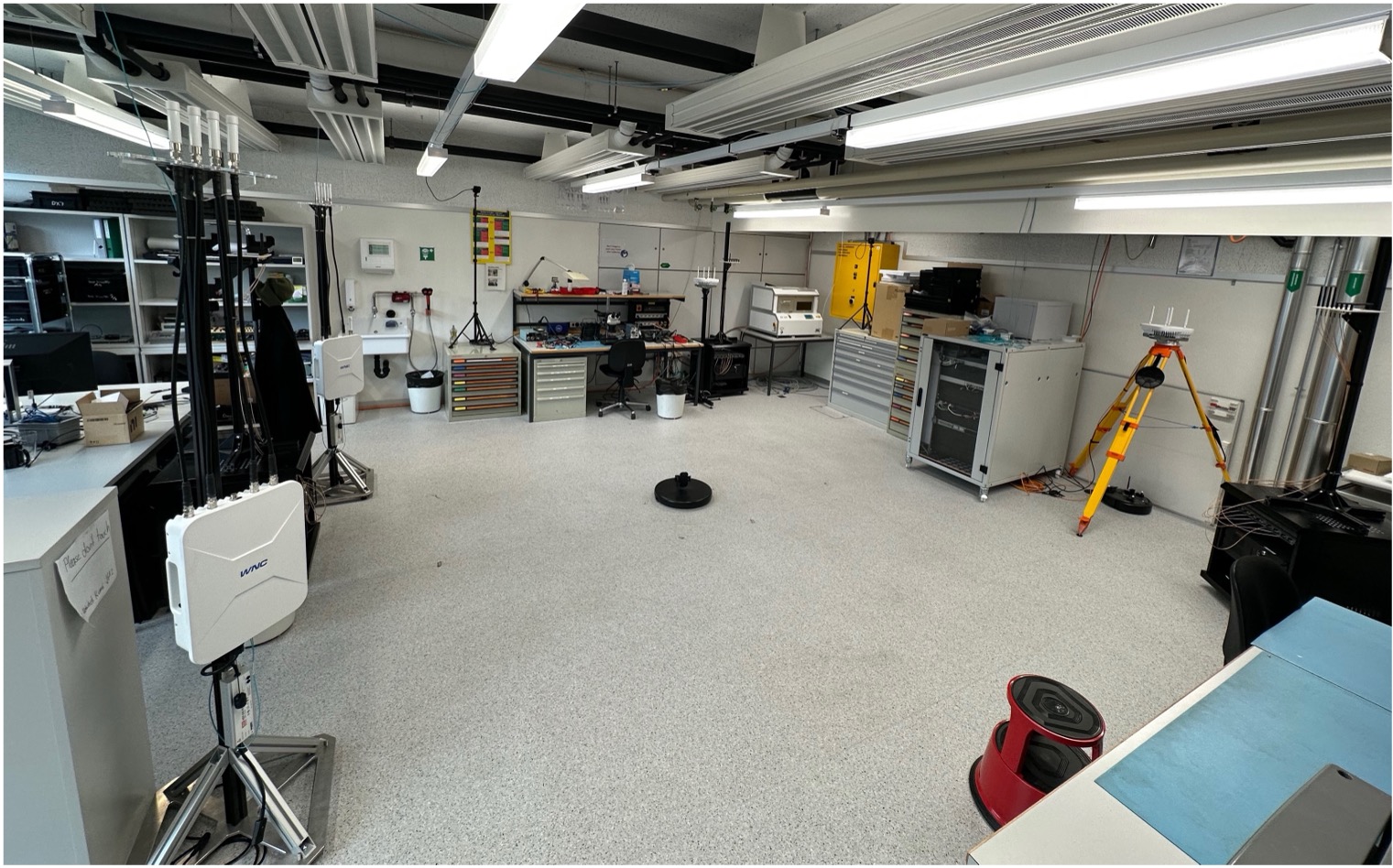}\label{fig:dev_class_measurement_room}%
  }\hfill
  \subfigure[]{
    \includegraphics[height=3.3cm]{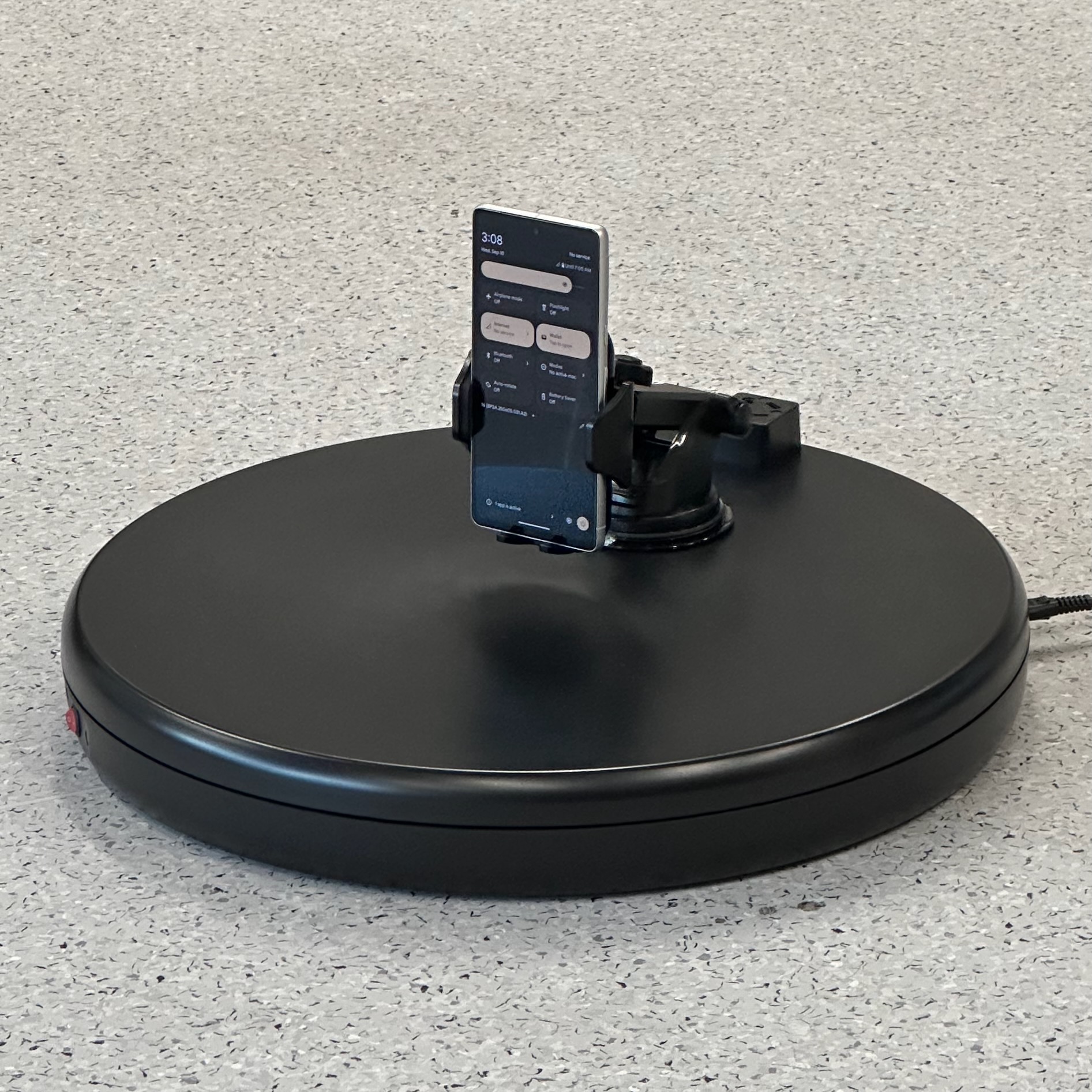}\label{fig:dev_class_rot_table}%
  }
  \subfigure[]{
    \includegraphics[width=0.99\columnwidth]{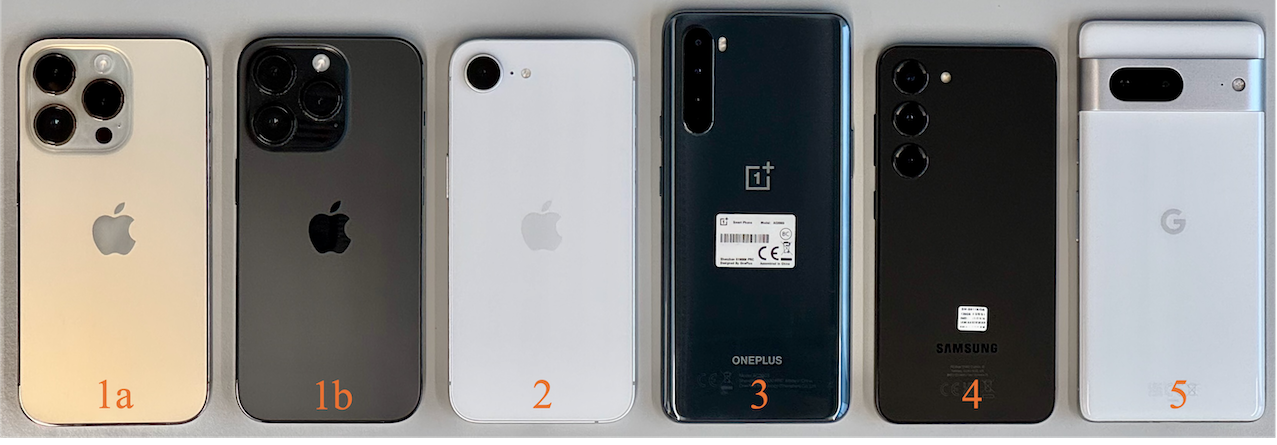}\label{fig:all_ues}%
  }
  \caption{Measurement setup for the \gls{CAEZ}-5G-DEV-CLASS dataset (a), rotation table with \gls{UE} mount (b), and the six measured \gls{COTS} \glspl{UE} (c).}
  \label{fig:caez_5g_dev_class}
\end{figure}

A Samsung Galaxy S23 (\gls{UE} 4 in \fref{fig:all_ues}) was mounted on a robot arm on top of a custom robot platform.
The robot was controlled manually.
Four WorldViz markers were mounted on the robot arm (which remained fixed) to enable tracking of the mounted \gls{UE}'s position and orientation.
Two measurement operators were present near the measurement area during \gls{CSI} collection.

\subsection{CAEZ-5G-DEV-CLASS}

\subsubsection{Overview}

The indoor measurement setup is shown in \fref{fig:dev_class_measurement_room}.
The setup is similar to that of CAEZ-5G-INDOOR, but the measurements were carried out on different days and no WorldViz PPT cameras were used. 
The measurement area is a joint lab/office space of about 4\,m\,$\times$\,4\,m between the lab desks.
The \glspl{ORU} were placed at the corners of the measurement area.
The dataset consists of six consecutive measurements, each of which is taken separately using one after the other \gls{UE} shown in \fref{fig:all_ues}.
The measurement protocol for each \gls{UE} comprises the following four steps: (i) rotation on a predefined fixed location with a rotation table, (ii) random human walk, (iii) additional rotations on the same location, and (iv) another random human walk on the next day.

\subsubsection{Measurement Details}

On the first measurement day, each measurement started when the \gls{UE} was mounted on the rotation table and taken out of flight mode.
As soon as the \gls{UE} connected to the 5G network, it was left slowly rotating for 30\,s.
Then, the operator removed the \gls{UE} from the phone mount and carried it randomly through the lab space for 60\,s.
Finally, the \gls{UE} was put back into the phone mount on the rotation table and left spinning for another 30\,s.
The measurement operator was present in the lab/office at all times.

On the second measurement day (one day after the first one), each measurement consisted only of 30\,s random human walk through the lab space.
This next-day dataset is used for testing.
Note that one of the \gls{ORU}'s power supplies stopped working overnight. Therefore, one \gls{ORU} was using a different power supply and was wired differently on the second day.
Also, the lab environment has slightly changed, as some chairs and equipment were moved from one day to the next. Therefore, this next-day evaluation dataset is not only recorded from different \gls{UE} locations, but also in a slightly modified environment.

%%%
\section{Case Studies of CSI-Based Sensing}

We now provide the implementation details and summarize experimental results for the three \gls{CSI}-based sensing tasks: (i) neural \gls{UE} positioning, (ii) channel charting, and (iii) closed-set device classification.
These experiments demonstrate the quality and utility of the measured real-world 5G NR \gls{CSI} datasets using state-of-the-art sensing algorithms.

\subsection{Machine-Learning Pipeline}

\begin{figure}[t]
    \centering
    \includegraphics[width=0.95\columnwidth]{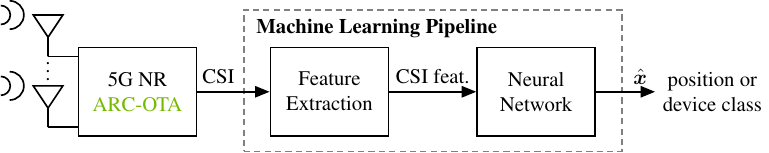}
    \caption{5G NR testbed and CSI-based machine-learning pipeline.}%
    \label{fig:pos_cc_pipeline}
\end{figure}

For the three studied sensing applications, we utilize a machine-learning pipeline as illustrated in \fref{fig:pos_cc_pipeline}, which depicts the high-level flow from the \gls{ARC-OTA} system to \gls{CSI} extraction, feature extraction, and \gls{NN} processing.

The implementation proceeds as follows.
The NVIDIA DataLake database holds \gls{FH} I/Q samples and L2 protocol information (FAPI), which enables offline processing of \gls{PUSCH} slots.
We use NVIDIA PyAerial, which provides Python bindings for the Aerial CUDA pipeline.
We run the PyAerial \gls{PUSCH} receiver up to the channel estimator to estimate \gls{CSI} for all four \glspl{ORU}.
The \gls{CSI} is estimated using PyAerial's default multi-stage minimum mean-square error channel estimator with time-delay estimation.
We save the \gls{CSI} estimates to disk, paired with the \gls{RNTI}, noise-variance estimates, and \gls{PUSCH} slot timestamps.

In a subsequent feature-extraction script, we read and process all \gls{CSI} estimates and the WorldViz position labels.
We linearly interpolate the WorldViz position labels to the timestamps of the \gls{PUSCH} slots.
The result of this data processing script is a \gls{CSI} feature array with timestamps and interpolated ground-truth position labels.
This data is then fed to the neural positioning or channel charting pipelines from \cite{zumegen2024software} and \cite{taner2025channel}, respectively.
For device classification, we do not collect ground-truth \gls{UE} positions and extract \gls{CSI} features directly in the device classification pipeline described in \fref{sec:dev_class}.

\subsection{CSI-based Neural \gls{UE} Positioning}

\subsubsection{Overview}

We apply the method from \cite{gonultacs2021csi} for neural \gls{UE} positioning.
The goal is to train a \gls{NN} that predicts absolute \gls{UE} position based on one \gls{CSI} feature; this is achieved through supervised training using ground-truth position labels.

\subsubsection{Neural Positioning Pipeline}

As \gls{CSI} features, we use down-sampled \gls{OFDM}-domain \gls{CSI} absolute values.
These are computed as follows.
We use full-spectrum \gls{CSI} estimates of all 273 \glspl{PRB} and compute their absolute values.
Low-pass filtering and down-sampling of the subcarrier dimension by a factor of 12 reduces the feature dimensions.
We average over all three \gls{DMRS} symbols in the same \gls{PUSCH} slot.
This results in 273 real-valued features per \gls{ORU} antenna.
The features from all \glspl{ORU} and all antennas are aggregated into one vector, which is then scaled to unit-norm.

The \gls{NN} architecture is a simple, fully-connected (feedforward) multi-layer perceptron with a probability map output, as used in~\cite{gonultacs2021csi}.\footnote{The \gls{NN} outputs a probability estimate for a set of predefined grid points: the so-called probability map~\cite{gonultacs2021csi}. We obtain the position estimate by computing the posterior-mean estimate based on the grid point positions and the associated probability estimates. This \gls{NN} architecture also enables us to compute a positioning variance estimate.}
Correspondingly, the ground-truth position labels are first mapped to ground-truth probability maps in a pre-processing step prior to \gls{NN} training.
The \gls{NN} is trained with a binary cross-entropy loss using mean-reduction computed between the ground-truth probability map points and the \gls{NN} output probability map points.

The positioning \gls{NN} is trained for 50 epochs with an initial learning rate of $10^{-4}$ and a batch size of 10 samples.
An epoch corresponds to a full pass over the entire training dataset.
We use the Adam optimizer in combination with a learning rate scheduler applying a step size decay of a factor $0.1$ after every~20 epochs.

The training and testing datasets are obtained by randomly partitioning the \gls{CAEZ} dataset into training and testing samples, applying a split of 80\,\% to 20\,\%, respectively.
The last 500 samples, which correspond to a single, connected sub-trajectory, are excluded from random partitioning and only taken for testing.
These samples provide insight into the generalization capabilities of the \gls{NN} as there are no neighboring training samples that are close in space and time.

\subsubsection{Results}

\begin{figure}[t]
  \centering
  \subfigure[Ground truth]{
    \includegraphics[width=0.45\columnwidth]{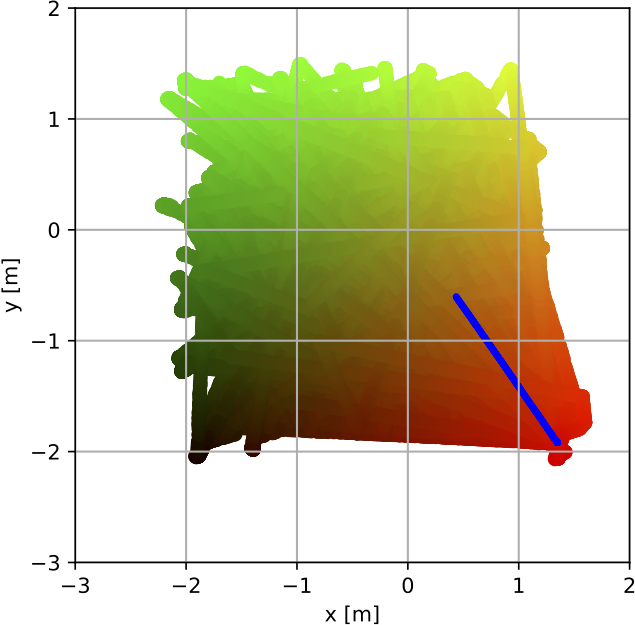}%
  }\hfill
  \subfigure[Estimate]{
    \includegraphics[width=0.45\columnwidth]{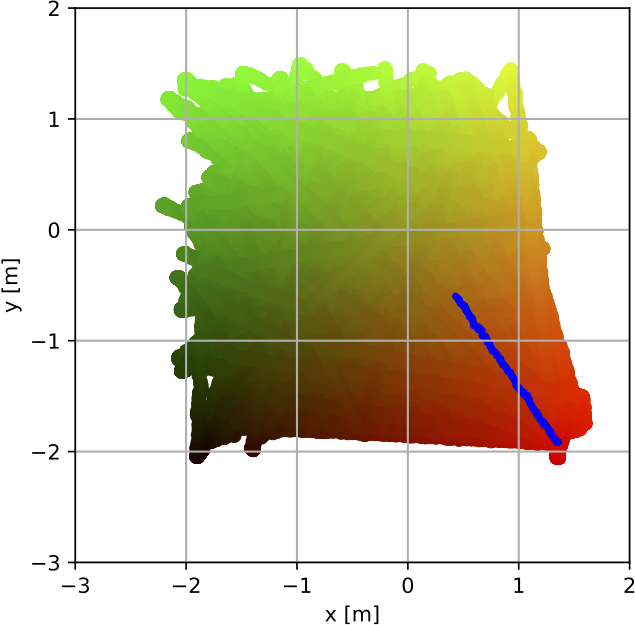}
  }
  \caption{Neural \gls{UE} positioning with CAEZ-5G-INDOOR dataset.}
  \label{fig:ue_pos_indoor}
\end{figure}

\fref{fig:ue_pos_indoor} shows neural \gls{UE} positioning results obtained with the CAEZ-5G-INDOOR dataset. The data points highlighted with a color gradient correspond to the randomly partitioned testing dataset. The blue trajectory corresponds to the last 500 data samples that are used for testing only. The randomly partitioned testing data achieves a mean absolute error of 0.6\,cm, and the last 500 test data points achieve a mean absolute error of~0.7\,cm.

\begin{figure}[t]
  \centering
  \subfigure[Ground truth]{
    \includegraphics[width=0.45\columnwidth]{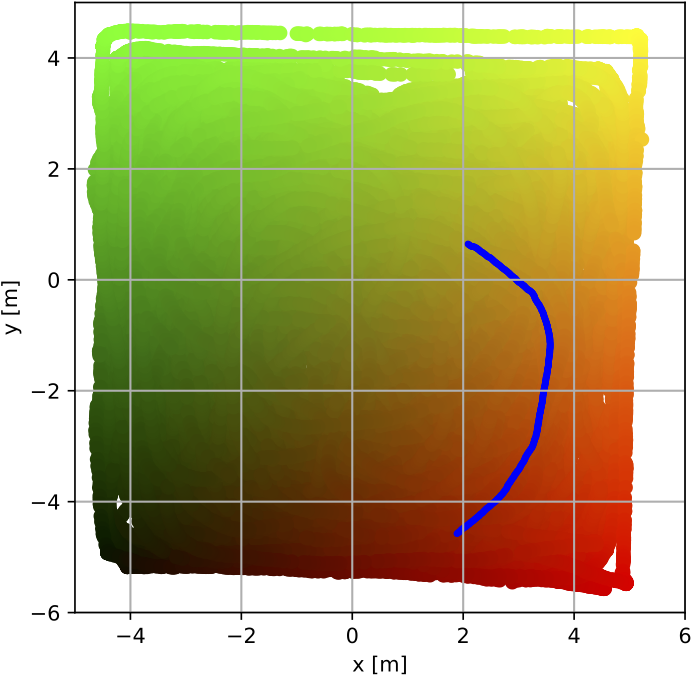}%
  }\hfill
  \subfigure[Estimate]{
    \includegraphics[width=0.45\columnwidth]{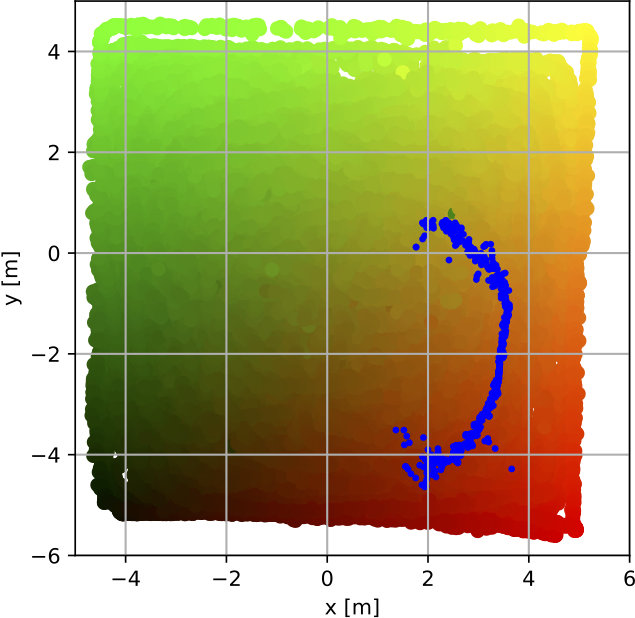}
  }
  \caption{Neural \gls{UE} positioning with CAEZ-5G-OUTDOOR dataset.}
  \label{fig:ue_pos_outdoor}
\end{figure}

\fref{fig:ue_pos_outdoor} shows neural \gls{UE} positioning results obtained with the CAEZ-5G-OUTDOOR dataset.
The test data achieves a mean absolute error of 5.7\,cm, and the last 500 test data points achieve a mean absolute error of 10\,cm.

\fref{tbl:abs_error} lists the mean, median, and 95th\,\% error of the randomly partitioned data points from all indoor and outdoor neural positioning and real-world channel charting experiments.
The high positioning accuracies achieved in all experiments demonstrate the utility of both CAEZ-5G-INDOOR and CAEZ-5G-OUTDOOR datasets for neural positioning in 5G.

\begin{table}[t]
\centering
\caption{Absolute positioning error [\cm].}
\label{tbl:abs_error}
\renewcommand{\arraystretch}{1.1}
\begin{tabular}{@{}p{4.6cm}ccc@{}}
\toprule
\textbf{CAEZ Dataset} & \textbf{Mean} & \textbf{Median} & \textbf{$95$th\,\%} \\ \midrule 
\multicolumn{4}{c}{\textit{ CSI-based Neural UE Positioning \cite{gonultacs2021csi} }} \\[0.75pt] 
\midrule
CAEZ-5G-INDOOR         & 0.6 & 0.5 & 1.3 \\
CAEZ-5G-OUTDOOR    & 5.7 & 4.6 & 13.2 \\ \midrule
\multicolumn{4}{c}{\textit{ Channel Charting in Real-World Coordinates \cite{taner2025channel} }} \\[0.75pt] \midrule
CAEZ-5G-OUTDOOR         & 73 & 64 & 158 \\ 
\bottomrule
\end{tabular}
\end{table}

\subsection{Channel Charting}

\subsubsection{Overview}

We apply the methods from \cite{ferrand2021triplet} and \cite{taner2025channel} for triplet-based channel charting and channel charting in real-world coordinates, respectively.
The goal is to train a \gls{NN} to predict \gls{UE} position based on a \gls{CSI} feature without ground-truth position knowledge.
This is achieved by dimensionality reduction techniques that preserve local proximity, i.e., points that are close in space will also be mapped closely in a low-dimensional pseudo space (i.e., the channel chart).
We train a \gls{NN} to learn this dimensionality reduction function that maps \gls{CSI} features to a two-dimensional representation.

We apply two channel charting techniques: (i) triplet-based channel charting \cite{ferrand2021triplet} and (ii) channel charting in real-world coordinates \cite{taner2025channel}.
The first method applies a triplet loss that is computed as follows.
For each \gls{CSI} sample (called anchor point), we take one sample that is close in time and another one that is far in time (but not too far).
These three samples constitute a set of three, i.e., a triplet.
The triplet loss then ensures that the distance in the channel chart between the close sample and the anchor is smaller than that of the anchor and the far sample.
The resulting channel chart typically preserves local closeness, but it is not bound to any real-world coordinates.

In order to ground the channel chart in real-world coordinates, we add the bilateration loss from \cite{taner2025channel} to the triplet loss.
This loss requires knowledge of the \glspl{ORU}' positions, which are typically known to the network operator.
The bilateration loss is computed from pairs of receive signal power estimates that can be simply computed from the \gls{CSI}.
The idea is that if the receive power at \gls{ORU} $b$ is much larger than the receive power at \gls{ORU} $b^\prime$, that \gls{CSI} sample should be mapped closer to the position of \gls{ORU} $b$ than to the position of \gls{ORU} $b^\prime$.

\subsubsection{Channel Charting Pipeline}

We use approximate autocorrelation \gls{CSI} features in the delay domain.
These are computed as follows.
We use full-spectrum \gls{CSI} estimates of all 273 \glspl{PRB} and compute their squared absolutes.
We average over all three \gls{DMRS} symbols in the same \gls{PUSCH} slot.
Then, we apply the inverse fast Fourier transform to obtain delay domain \gls{CSI} features.
We truncate the feature vector to the first 25 complex-valued taps.
Stacking the real and imaginary parts yields 50 real-valued features per \gls{ORU} antenna.
Finally, we aggregate and vectorize the resulting features from all \glspl{ORU} and all antennas, and normalize the vector to unit-norm.

The \gls{NN} architecture is a fully-connected multi-layer perceptron with a linear output layer, as implemented in \cite{taner2025channel}.
The channel charting \gls{NN} is trained for 300 epochs (200 for real-world channel charting) with an initial learning rate of $10^{-3}$ and a batch size of 100 samples (256 for real-world channel charting).
Training uses one triplet per anchor (two for real-world channel charting).
We use the Adam optimizer combined with a learning rate scheduler that applies a step size decay of a factor $0.1$ after every 200 epochs (50 epochs for real-world channel charting).
For the bilateration loss, we select a receive power margin of 13\,dB.

The training and testing datasets are obtained by randomly partitioning the \gls{CAEZ} dataset into training and testing samples, applying a split of 80\,\% to 20\,\%, respectively.
Again, the last 500 samples are excluded from random partitioning and reserved for additional testing.

\subsubsection{Results}

\begin{figure*}[t]
  \centering
  \subfigure[Ground truth]{
    \includegraphics[width=0.3\textwidth]{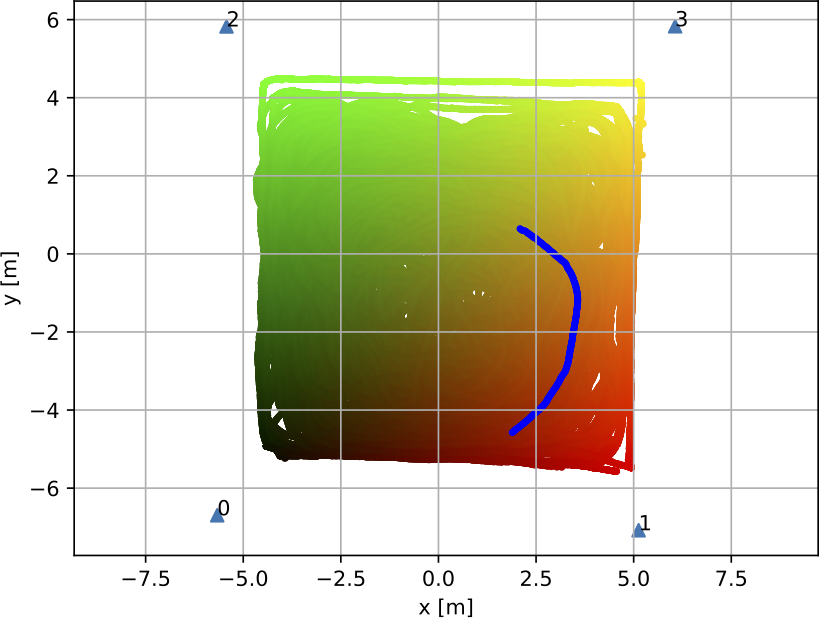}\label{fig:cc_outdoor_gt}%
  }\hfill
  \subfigure[Channel chart trained with triplet loss]{
    \includegraphics[width=0.3\textwidth]{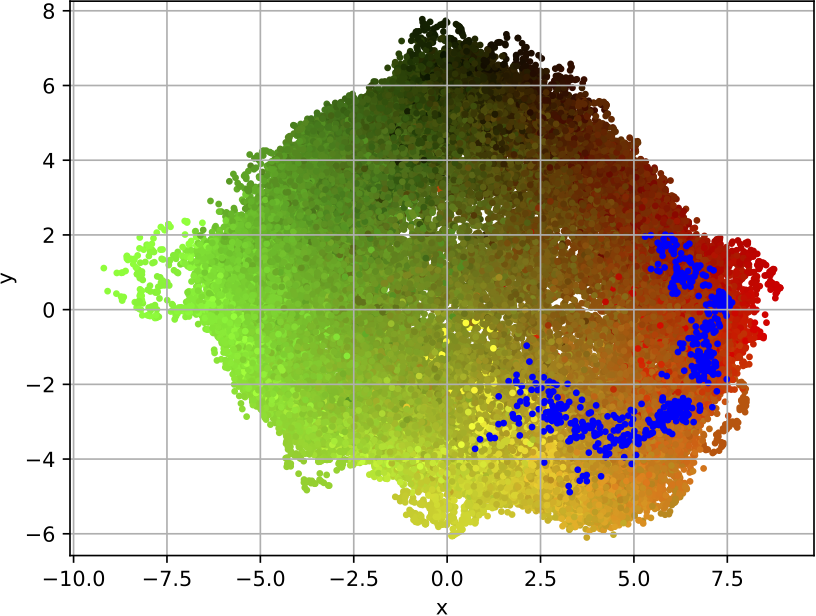}\label{fig:cc_triplet}
  }\hfill
  \subfigure[Channel chart in real-world coordinates]{
    \includegraphics[width=0.3\textwidth]{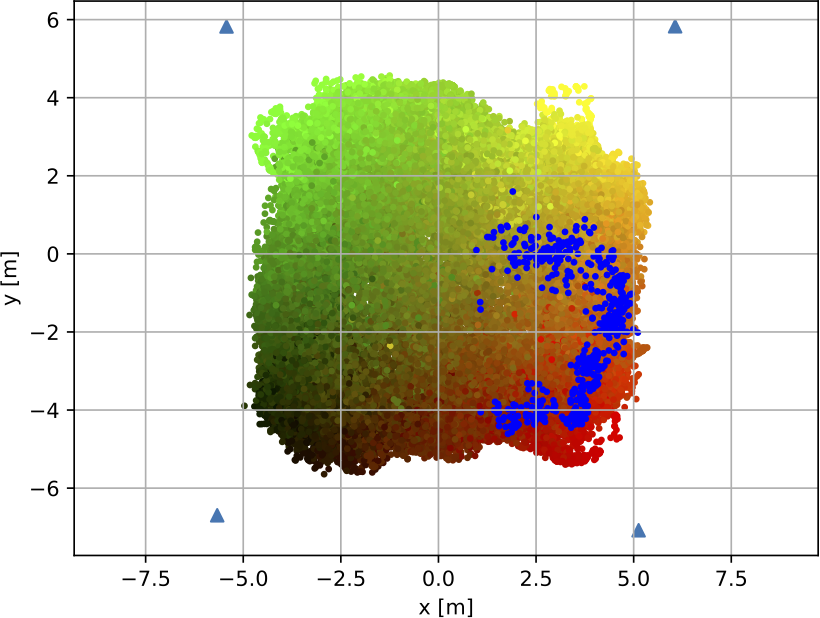}\label{fig:cc_real-world}
  }
  \caption{Channel charting with CAEZ-5G-OUTDOOR dataset. (a) shows the ground-truth locations of the test data points and the \gls{ORU} locations (blue triangles), (b) shows the channel chart obtained with the triplet loss \cite{ferrand2021triplet}, and (c) shows the channel chart in real-world coordinates \cite{taner2025channel}.}
  \label{fig:cc_outdoor}
\end{figure*}

\fref{fig:cc_outdoor} shows channel charting results with the CAEZ-5G-OUTDOOR dataset.
The data points shaded with a color gradient correspond to the randomly partitioned testing dataset.
The blue trajectory denotes the last 500 data samples that are used for testing only.

\fref{fig:cc_triplet} shows the channel chart after training the \gls{NN} using only the triplet loss.
This channel chart achieves a continuity of 98.2\,\% and a trustworthiness of 97.4\,\%.

\fref{fig:cc_real-world} shows the channel chart in real-world coordinates after training the \gls{NN} with both the triplet and the bilateration loss (and a bounding box loss not further described in this paper).
This result shows that the bilateration loss is sufficient to ground the channel chart in real-world coordinates.
This also enables us to measure the mean absolute positioning error, which is 73\,cm.
The continuity marginally decreases to 98.0\,\%, and the trustworthiness remains at 97.4\,\%.
This result demonstrates that absolute \gls{UE} positioning is possible with channel charting in a 5G NR system. 

\subsection{CSI-based Device Classification}\label{sec:dev_class}

\subsubsection{Overview}

We build on the closed-set classification pipeline published in the simulation code of \cite{shen2022towards}.
We train a \gls{NN} that identifies a \gls{UE} within a closed set of candidates based on the \gls{RF} fingerprint.
Since the measured \gls{OFDM}-domain \gls{CSI} contains the product of the transmitter's \gls{RF} chain impulse response (which is part of the fingerprint) and the physical \gls{RF} channel between the \gls{UE} and the \glspl{ORU}, we need to extract \gls{RFFI} features that are independent of the transmitter's location (i.e., independent from the wireless channel).
To this end, we extract \gls{RFFI} features inspired by the \gls{CSI} obfuscation method proposed in \cite{stephan2025csi}.
As a result, we can use supervised \gls{NN} training to learn an accurate prediction of the \gls{UE} type (i.e., the device class) from \gls{CSI}-based \gls{RFFI} features, independently of the transmitter's location and \gls{RF} environment.

\subsubsection{Device Classification Pipeline}

We compute \gls{CSI} features as follows.
For each of the three \glspl{DMRS} in the same \gls{PUSCH} slot, we gather the \gls{CSI} of all four \glspl{ORU} with four antennas each in the columns of the channel matrix $\bH\in\complexset^{3276\times16}$.
We then stack the \gls{CSI} matrices from all three \glspl{DMRS} along the subcarrier dimension (i.e., rows) and normalize all columns to unit norm.
We then compute the compact singular value decomposition of the resulting matrix and take the dominant left singular vector.
Intuitively, this vector contains, for each subcarrier and each \gls{DMRS}, the common part across all distributed receive antennas that is mostly influenced by the transmit \gls{RF} circuitry---but not by the wireless channel. 
We then reshape the singular vector to recover the subcarrier and time dimensions, and stack the real and imaginary parts in the last tensor dimension and obtain our \gls{RFFI} feature $\bmf\in\reals^{3276\times3\times2}$.
This feature extraction process is implemented for each \gls{PUSCH} slot (i.e., for each \gls{CSI} sample) independently.

We utilize the closed-set training pipeline from \cite{shen2022towards} and modify it to use our 5G \gls{RFFI} features.
The \gls{NN} architecture is a 2D convolutional ResNet model with a softmax output layer.
We train the \gls{NN} with a categorical cross-entropy loss on the one-hot ground-truth device class labels.
Each \gls{UE} shown in \fref{fig:all_ues} is associated with an individual device class.

The classification \gls{NN} is trained for at most 400 epochs with an initial learning rate of $10^{-3}$ and a batch size of 32 samples.
We use the RMSprop optimizer in combination with a learning rate scheduler that applies a step size decay of a factor $0.2$ after~10 consecutive epochs without improvement.
Early stopping terminates the training process after 30 consecutive epochs without improvement.
The learning rate scheduler and the early stopping mechanism operate on the validation loss computed from the same-day testing dataset described below.

We apply two testing datasets.
For the same-day testing dataset, for each \gls{UE}, we first sort the samples with respect to the measurement timestamp and take 12.5\,\% of the samples from the center.
In our measurement protocol, this corresponds to a fraction of the time when the \gls{UE} was randomly moved through the lab space.
The remaining 87.5\,\% of sample points are taken for training.
The next-day measurement campaign is taken solely for testing purposes.

\subsubsection{Results}

\begin{figure}[htp]
    \centering
    \subfigure[NN trained without UE 1b]{\includegraphics[scale=0.65]{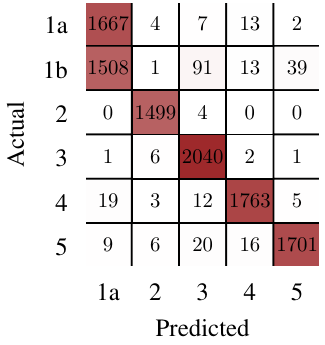}%
    \label{fig:device_class_same_day_5}}
    \hfill
    \subfigure[NN trained with all UEs]{\includegraphics[scale=0.65]{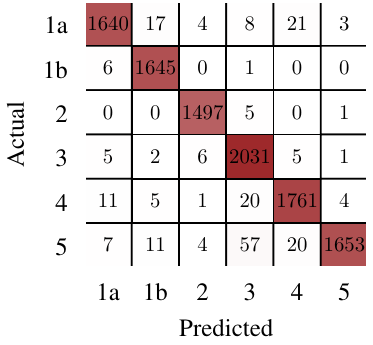}%
    \label{fig:device_class_same_day_6}}
    \caption{Device classification same-day evaluation: (a) NN trained without UE 1b achieves 99\% accuracy and (b) NN trained with all UEs achieves 98\% accuracy. UE 1a and UE 1b are both an Apple iPhone 14 Pro.}%
    \label{fig:device_class_same_day}
\end{figure}

\begin{figure}[htp]
    \centering
    \subfigure[NN trained without UE 1b]{\includegraphics[scale=0.65]{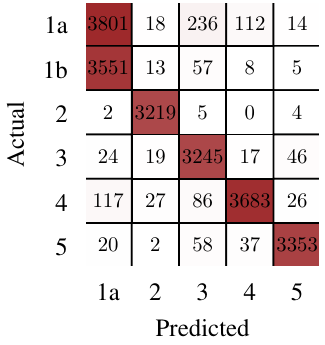}%
    \label{fig:device_class_next_day_5}}
    \hfill
    \subfigure[NN trained with all UEs]{\includegraphics[scale=0.65]{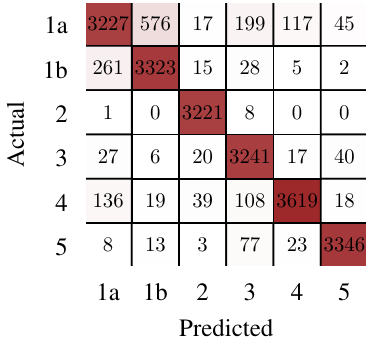}%
    \label{fig:device_class_next_day_6}}
    \caption{Device classification next-day evaluation: (a) NN trained without UE 1b achieves 95\% accuracy and (b) NN trained with all UEs achieves 92\% accuracy. UE 1a and UE 1b are both an Apple iPhone 14 Pro.}%
    \label{fig:device_class_next_day}
\end{figure}

\fref{fig:device_class_same_day} and \fref{fig:device_class_next_day} show two different experiments evaluated on the same day and on the next day, respectively.
For the results shown in \fref{fig:device_class_same_day_5} and \fref{fig:device_class_next_day_5}, the classification \gls{NN} was trained without \gls{UE} 1b.
For the results shown in \fref{fig:device_class_same_day_6} and \fref{fig:device_class_next_day_6}, the classification \gls{NN} was trained with all \glspl{UE} shown in \fref{fig:all_ues}.
Since \gls{UE} 1a and \gls{UE} 1b are both an Apple iPhone 14 Pro, this experiment highlights whether an unknown \gls{UE} (i.e., 1b) gets classified as iPhone 14 Pro (i.e., as 1a), and whether it is possible to distinguish two \glspl{UE} of the same model (i.e., 1a and 1b).

In the same-day evaluation shown in \fref{fig:device_class_same_day_5}, the \gls{NN} achieves a classification accuracy of 99\,\% within the five \glspl{UE} for which the \gls{NN} is trained.
Interestingly, 91\,\% of \gls{UE} 1b's test samples are classified as 1a.
In the same-day evaluation shown in \fref{fig:device_class_same_day_6}, the \gls{NN} achieves an overall accuracy of 98\,\% across all \glspl{UE} shown in \fref{fig:all_ues}.

In the next-day evaluation shown in \fref{fig:device_class_next_day_5}, the classification accuracy slightly decreases to 95\,\% within the five \glspl{UE} for which the \gls{NN} is trained.
However, 98\,\% of \gls{UE} 1b's test samples are classified as \gls{UE} 1a.
In the next-day evaluation shown in \fref{fig:device_class_next_day_6}, the \gls{NN} achieves an overall accuracy of 92\,\% for all \glspl{UE} shown in \fref{fig:all_ues}.
Interestingly, most confusions occur between \gls{UE} 1a and \gls{UE} 1b, which are both the same model (Apple iPhone 14 Pro).

All of these results demonstrate that accurate closed-set device classification is possible in a 5G NR system. 

\section{Conclusions}

We have published the first real-world 5G NR \gls{CSI} datasets with position labels and complex-valued full-spectrum \gls{CSI} samples.
The datasets CAEZ-5G-INDOOR and CAEZ-5G-OUTDOOR are position-tagged and enable high-accuracy neural \gls{UE} positioning and channel charting in real-world coordinates.
The CAEZ-5G-DEV-CLASS dataset enables highly accurate device classification with location-independent \gls{RFFI} features and testing across two days with slightly modified conditions.
Our experimental results demonstrate high accuracy for all three tasks: (i) neural \gls{UE} positioning with mean absolute errors of 0.6\,cm (indoor) and 5.7\,cm (outdoor), (ii) channel charting in real-world coordinates with 73\,cm mean absolute error, and (iii) device classification with accuracies of 99\,\% (same-day) and 95\,\% (next-day).
The \gls{CAEZ} datasets and simulation code are publicly available at \url{https://caez.ethz.ch}.

Future work includes collecting more datasets, including mixed \gls{LOS} and non-\gls{LOS} scenarios, larger measurement areas, and three-dimensional (non-planar) \gls{UE} trajectories.
We also plan to study real-world validation of model-based receivers \cite{Wiesmayr2022} and \gls{NN}-based receivers \cite{wiesmayr2025design}.

\linespread{0.985}

\bibliographystyle{IEEEtran}

\end{document}